\documentclass[fleqn,useAMS]{Papers}
\usepackage{newtxtext,newtxmath}
\usepackage[T1]{fontenc}
\usepackage{subfloat}
\usepackage{makecell}

\usepackage[square,sort,numbers]{natbib}
\usepackage{graphicx}	
\usepackage{amsmath}	
\usepackage{lipsum}
\usepackage{caption}
\usepackage{comment}
\usepackage{caption}
\usepackage{lineno}
\newcommand{\rev}[1]{{\color{black} #1}} 

\title[Theory for halo mass functions and density profiles]{Dark matter halo mass functions and density profiles from mass and energy cascade}

\author[Z. Xu]{Zhijie (Jay) Xu,$^{1}$\thanks{E-mail: \href{mailto:zhijie.xu@pnnl.gov}{zhijie.xu@pnnl.gov};} 
\\
$^{1}$Physical and Computational Sciences Directorate, Pacific Northwest National Laboratory; Richland, WA 99354, USA\\
}

\date{Accepted XXX. Received YYY; in original form ZZZ}

\pubyear{2023}

\begin{document}
\label{firstpage}
\pagerange{\pageref{firstpage}--\pageref{lastpage}}
\maketitle

\begin{abstract}
Halo abundance and structure play a central role for modeling structure formation and evolution. Without relying on a spherical or ellipsoidal collapse model, we analytically derive the halo mass function and cuspy halo density (inner slope of -4/3) based on the mass and energy cascade theory in dark matter flow. The hierarchical halo structure formation leads to halo or particle random walk with a position-dependent waiting time $\tau_g$. First, the inverse mass cascade from small to large scales leads to the halo random walk in mass space with $\tau_g\propto m_h^{-\lambda}$, where $m_h$ is the halo mass and $\lambda$ is a halo geometry parameter with predicted value of 2/3. The corresponding Fokker-Planck solution for halo random walk in mass space gives rise to the halo mass function with a power-law behavior on small scale and exponential decay on large scale. This can be further improved by considering two different $\lambda$ for haloes below and above a critical mass scale $m_h^*$, i.e. a double-$\lambda$ halo mass function. Second, a double-$\gamma$ density profile can be derived based on the particle random walk in 3D space with a position-dependent waiting time $\tau_g \propto \Phi(r)^{-1} \propto r^{-\gamma}$, where $\Phi$ is the gravitational potential and $r$ is the particle distance to halo center. Theory predicts $\gamma=2/3$ that leads to a cuspy density profile with an inner slope of -4/3, consistent with the predicted scaling laws from energy cascade. The Press-Schechter mass function and Einasto density profile are just special cases of proposed models. The small scale permanence can be identified due to the scale-independent rates of mass and energy cascade, where density profiles of different halo masses and redshifts converge to the $-4/3$ scaling law ($\rho_h \propto r^{-4/3}$) on small scales. Theory predicts the halo number density scales with halo mass as $\propto m_h^{-1.9}$, while the halo mass density scales as $\propto m_h^{4/9}$. Results were compared against the Illustris simulations. This new perspective provides a theory for nearly universal halo mass functions and density profiles.
\end{abstract}

\begin{keywords}
\vspace*{-10pt}
Dark matter halo; Mass function; Density profile; Random walk; 
\end{keywords}

\begingroup
\let\clearpage\relax
\tableofcontents
\endgroup
\vspace*{-20pt}

\section{Introduction}
\label{sec:1}
Within the standard $\Lambda$CDM (cold dark matter) cosmology \citep{Peebles:1984-Tests-of-cosmological-models,Spergel:2003-First-Year-Wilkinson-Microwave-Anisotropy,Komatsu:Seven-year-Wilkinson-Microwave-Anisotropy-Probe,Frenk:2012-Dark-matter-and-cosmic-structure}, the formation of structures proceeds hierarchically with small structures coalescing into large structures in a "bottom-up" fashion. For systems involving long-range interaction, the formation of haloes of different sizes is necessary to maximize system entropy \citep{Xu:2023-Maximum-entropy-distributions-of-dark-matter}. Therefore, highly localized halo structures and their evolution are major features of $\Lambda$CDM model \citep{Neyman:1952-A-Theory-of-the-Spatial-Distri,Cooray:2002-Halo-models-of-large-scale-str}. As a counterpart of "eddies" in hydrodynamic turbulence, "haloes" are the building blocks in the flow of dark matter \citep{Xu:2022-Dark_matter-flow-and-hydrodynamic-turbulence-presentation, Xu:2023-Universal-scaling-laws-and-density-slope, Xu:2021-Inverse-mass-cascade-mass-function}. 
Halo abundance and internal structure play a central role for modeling structure formation and evolution. These two quantities are also critical to understand the small scale challenges for $\Lambda$CDM when comparing model with observations \citep{Flores:1994-Observational-and-Theoretical-Constraints,deBlok:2009-The-Core-Cusp-Problem,Klypin:1999-Where-Are-the-Missing-Galactic,Boylan_Kolchin:2011-Too-big-to-fail}. However, despite having been extensively studied over many decades, our understanding is still not entirely satisfactory. 

First, the abundance of dark matter haloes is described by a halo mass function. The seminal Press-Schechter (PS) model allows one to predict the shape and evolution of mass function based on a density peak approach \citep{Press:1974-Formation-of-Galaxies-and-Clus}. This model relies on a threshold value of overdensity ($\delta_c$) that can be obtained from the nonlinear collapse of a spherical over-density \citep{Tomita:1969-Formation-of-Gravitationally-B,Gunn:1972-Infall-of-Matter-into-Clusters}. Bond et al. provided an alternative derivation using an excursion set approach (EPS) that puts the theory on a firmer footing by removing the fudge factor in original PS model \citep{Bond:1991-Excursion-Set-Mass-Functions-f}, which was further extend to excursion set with correlated steps \citep{Musso:2012-One-step-beyond-the-excursion-set-approach,Paranjape:2012-Peaks-theory-and-the-excursion-set-approach,Maggiore:2010-THE-HALO-MASS-FUNCTION-FROM-EXCURSION-SET-THEORY}. The PS model was further improved by Jedamzik with a formalism explicitly counting all cosmic materials to address the so-called "cloud-in-cloud" problem in density peak approach \citep{Jedamzik:1995-The-Cloud-in-Cloud-Problem-in-the-Press-SchechterJ}. Lee and Shandarin adopted Zeldovich approximation and extended the PS formalism to a non-spherical dynamical model \citep{Lee:1998-The-Cosmological-Mass-Distribution-Function}. Other developments include combination of the peak and excursion set approaches \citep{Paranjape:2012-Peaks-theory-and-the-excursion-set-approach}, a moving barrier as a better density threshold \citep{Corasaniti:2011-Excursion-set-halo-mass-function-and-bias}, and more recent efforts on developing emulators of halo mass functions for a range of different cosmologies \citep{Bocquet:2020-The-Mira-Titan-Universe}. 

However, when compared to N-body simulations, both PS and EPS models overestimate the number of low-mass haloes and underestimate the number of massive haloes. There are also significant errors at high redshifts \citep{Springel:2005-Simulations-of-the-formation--}. Further improvement was achieved by computing the density threshold $\delta_c$ for ellipsoidal collapse \citep{Sheth:2001-Ellipsoidal-collapse-and-an-im,Sheth:1999-Large-scale-bias-and-the-peak-}. In contrast to the spherical collapse where $\delta_{c} $ is independent of halo mass, the ellipsoidal collapse leads to a mass-dependent overdensity threshold $\delta_c$. This modification (hereafter ST) considerably complicates the derivation but provides a better agreement with simulations.

Because of its simplicity, the PS-EPS-ST mass functions are still a very popular analytic model. However, the theoretical basis of this approach is at best heuristic. First, the derivation requires a threshold overdensity from a simplified (if not over simplified) collapse model (either spherical or ellipsoidal). Second, the linear density field is required to identify collapsed structures that is deeply in the non-linear regime. In principle, halo mass function should be an objective intrinsic property of self-gravitating collisionless system that is independent of any simplified  (spherical or ellipsoidal) collapse models. In this paper, a different approach is taken to derive the halo mass function without resorting to any simplified models. This approach is based on the random walk of haloes in mass space, which is a direct result of inverse mass cascade in dark matter flow \citep{Xu:2021-Inverse-mass-cascade-mass-function}. 

Next, the structure of haloes is described by the halo density profile 
that can be studied both analytically and numerically with \textit{N}-body simulations \citep{Moore:1998-Resolving-the-structure-of-col,Klypin:2001-Resolving-the-structure-of-col}. Since the seminal work of spherical collapse \citep{Gunn:1972-Infall-of-Matter-into-Clusters}, the power-law density profile was derived under the self-similar approximation. The secondary in-fall model suggests a power-law density dependent on the initial density of the region that collapsed \citep{Bertschinger:1985-Self-Similar-Secondary-Infall-,Fillmore:1984-Self-Similar-Gravitational-Col}. High-resolution \textit{N}-body simulations have shown nearly universal profile with a cuspy density shallower than isothermal profile at smaller radius and steeper at larger radius \citep{Navarro:1997-A-universal-density-profile-fr,Navarro:2004-The-inner-structure-of-ACDM-ha}. For the cuspy inner density from N-body simulations, there seems no consensus on the exact value of the asymptotic logarithmic density slope $\gamma$. Since the first prediction of $\gamma = -1.0$ in NFW profile \citep{Navarro:1997-A-universal-density-profile-fr}, the inner density slope of simulated haloes have different values from $\gamma>-1.0$ \citep{Navarro:2010-The-diversity-and-similarity-of-simulated} to $\gamma=-1.2$ \citep{Diemand:2011-The-Structure-and-Evolution-of-Cold-Dark}, and $\gamma \approx -1.3$ \citep{Governato:2010-Bulgeless-dwarf-galaxies-and-dark-matter-cores,McKeown:2022-Amplified-J-factors-in-the-Galactic-Centre,Lazar:2020-A-dark-matter-profile-to-model-diverse}. 
In addition, there still lacks a complete understanding for the origin of nearly universal density profile \citep{Cooray:2002-Halo-models-of-large-scale-str}. In this paper, similar to the halo random walk in mass space for halo mass function, a new approach is presented based on the particle random walk in real space, which provides a possible theory for nearly universal halo structures and density profiles.

\section{Existing halo mass functions}
For comparison with our mass function model, a brief overview of existing mass functions is presented here. The exact definition of mass function varies widely in the literature. The two widely used mass functions are defined as
\begin{equation} 
\label{eq:1-1} 
\begin{split}
&F_M(m_h,z) \equiv \frac{dn(m_h,z)}{d\ln(m_h)}\textrm{,} \quad \textbf{f}(\sigma_{\delta},z)\equiv F_M\frac{m_h}{\rho_0}\frac{d\ln(m_h)}{d\ln(\sigma_{\delta}^{-1})},         
\end{split}
\end{equation} 
where $n(m_h,z)$ is the number density of haloes, $\rho_0$ is the background density. Here $\sigma_{\delta} \left(m_{h} \right)$ is the density fluctuation when density field is smoothed at mass scale $m_{h}$, which can be computed from the density power spectrum. When a normalized variable $\nu ={\delta _{c}^{2} /\sigma _{\delta }^{2} \left(m_{h} \right)}$ is used, the third definition $f(\nu)$ can be introduced such that the multiplicity mass function $\textbf{f}(\sigma_{\delta},z)=2\nu f(\nu)$. In this definition, the PS mass function reads
\begin{equation} 
\label{eq:1} 
f_{PS} \left(\nu \right)=\frac{1}{\sqrt{2\pi } \sqrt{\nu } } e^{-{\nu /2} }.         
\end{equation} 
The modified PS model (ST model) can be compactly written as:
\begin{equation} 
\label{eq:2} 
f_{ST} \left(\nu \right)=A\sqrt{\frac{2q}{\pi } } \left(1+\frac{1}{\left(q\nu \right)^{p} } \right)\frac{1}{2\sqrt{\nu } } e^{-{q\nu /2} } ,       
\end{equation} 
where the normalization condition requires: 
\begin{equation} 
\label{eq:3} 
A=\frac{\sqrt{\pi } }{\Gamma \left({1/2} \right)+2^{-p} \Gamma \left({1/2} -p\right)} .         
\end{equation} 
The best fitted parameters from simulation is $A=0.3222$, $q=0.707$, and $p=0.3$ (hereafter ST1), while $A=0.3222$, $q=0.75$, and $p=0.3$ was suggested by \citet{Sheth:2002-An-excursion-set-model-of-hier} (hereafter ST2). Both models satisfy the normalization condition $\int _{0}^{\infty }f\left(\nu \right) d\nu =1$.

Many empirical mass functions were also proposed by fitting to the high-resolution simulation data. For example, a universal mass function by Jenkins etc. (hereafter JK) covers a wide range of different cosmologies and redshifts that is written as \citep{Jenkins:2001-The-mass-function-of-dark-matt},
\begin{equation} 
\label{eq:4} 
f_{JK} \left(\nu \right)=\frac{0.315}{2\nu } \exp [-\left|\ln \left({\sqrt{v} /\delta _{c} } \right)+0.61\right|^{3.8} ] ,       
\end{equation} 
where the threshold density $\delta _{c} =1.6865$. Using a similar form of mass function to ST, Warren proposed (hereafter WR) \citep{Warren:2006-Precision-determination-of-the}
\begin{equation} 
\label{eq:4-2} 
f_{WR} \left(\nu \right) = 0.7234\left[\left(\frac{\delta_c}{\sqrt{\nu}}\right)^{-1.625}+0.2538\right]\exp\left(-\frac{1.1982}{\delta_c^2/\nu}\right),   
\end{equation} 
It should be noted that these empirical mass functions might not satisfy the normalization constraint and can be difficult to extrapolate beyond the range of fit.

\rev{The other widely used empirical mass function by Tinker etc. was also calibrated from numerical simulations with haloes identified as isolated spherical overdensity masses. The range of halo mass is between $10^{11}$ and $10^{15}$ $h^{-1}M_{\odot}$ with redshift $z\le 2$ \citep{Tinker:2008-Toward-a-Halo-Mass-Function-for-Precision-Cosmology}. TK mass function reads 
\begin{equation}
\begin{split}
&\textbf{f}(\sigma_{\delta},z) = A\left[\left(\frac{\sigma_{\delta}}{b}\right)^{-a}+1\right]\exp\left[-\frac{c}{\sigma_{\delta}^2}\right],\\
&\textrm{or equivalently,}\\
&f_{TK}(\nu) = \frac{A}{2\nu}\left[\left(\frac{\delta_c}{b\sqrt{\nu}}\right)^{-a}+1\right]\exp\left[-\frac{c\nu}{\delta_c^2}\right],
\end{split}
\label{eq:4-50}
\end{equation}
where best fitted parameters $A=0.186$, $a=1.47$, $b=2.57$ and $c=1.19$ for haloes with a critical density ratio $\Delta_c=200$.} Table \ref{tab:1} summarizes different halo mass functions $\textbf{f}(\sigma_{\delta}, z)$ in Eq. \eqref{eq:1-1}. The double-$\lambda$ mass function is analytically derived in Section \ref{sec:4}. 

\begin{table*}
\begin{center}
\captionsetup{justification=centering}
\centering
\caption{Different Halo Mass Functions $\textbf{f}(\sigma_{\delta}, z)$}
\begin{tabular}{llcc}
\hline\hline
\qquad Reference  & \qquad Mass Function $\textbf{f}(\sigma_{\delta}, z)$& Mass Range of Fit &  Redshift range of Fit \\  \\

PS, Press \& Schechter  &   $\sqrt{\frac{2}{\pi}} \frac{\delta_c}{\sigma_{\delta}}\exp\left[-\frac{\delta_c^2}{2\sigma_{\delta}^2}\right]$     &unspecified     & unspecified   \\[3ex]

ST, Sheth \& Tormen & $A \sqrt{\frac{2q}{\pi}}
\frac{\delta_c}{\sigma_{\delta}} \exp  \left[ - \frac{q \delta^2_c}{2\sigma_{\delta}^2} \right] \left[ 1 + \left( \frac{\sigma_{\delta}^2}{q
      \delta^2_c} \right) ^ p \right]$ & unspecified & unspecified\\ [3ex]

JK, Jenkins et al. &$ 0.315 \exp \left[ -| \ln \sigma_{\delta}^{-1} + 0.61 |^{3.8} \right]$ & $-1.2\le\ln\sigma_{\delta}^{-1}\le1.05$ & $z=0-5$\\[3ex]

WR, Warren et al. & $0.7234 \left( \sigma_{\delta}^{-1.625} + 0.2538 \right)\exp \left[ -\frac{1.1982}{\sigma_{\delta}^2}\right]$ &
$(10^{10}-10^{15}) h^{-1}M_{\odot} $ & $z=0$\\[3ex]

TK, Tinker et al. & ${A}\exp\left[-\frac{c}{\sigma_{\delta}^2}\right]\left[\left(\frac{\sigma_{\delta}}{b}\right)^{-a}+1\right]$ &
$(10^{11}-10^{15}) h^{-1}M_{\odot} $ & $z=0-2$\\[4ex]

Double-$\lambda$, Xu (this work Eq. \eqref{eq:13}) & $\frac{2p(2\sqrt{\eta_{0} })^{-q} }{\Gamma \left({q/2} \right)}{(\frac{\delta_c}{\sigma_{\delta}}})^{pq} \exp\left[-\frac{1}{4\eta_0}\left(\frac{\delta_c}{\sigma_{\delta}}\right)^{2p}  \right]$ & unspecified & unspecified\\ [4ex]
      
\hline\hline

\label{tab:1} 
\end{tabular}
\end{center}
\end{table*}

\section{Mass and energy cascade between haloes}
\begin{figure}
\includegraphics*[width=\columnwidth]{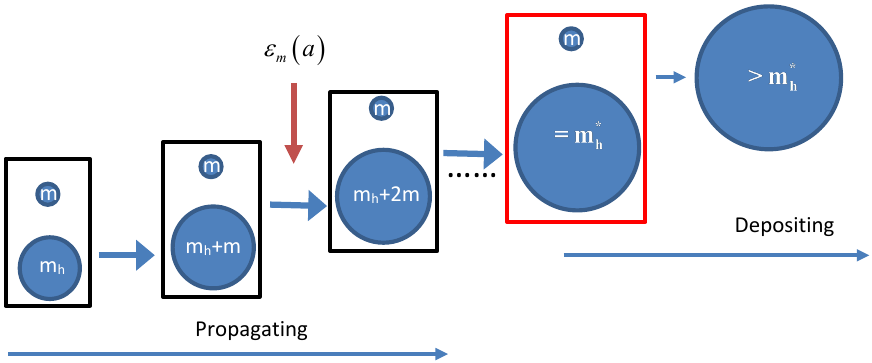}
\caption{Schematic plot of the inverse mass cascade for hierarchical structure formation. Halo of mass $m_h$ merges with single merger (free DM particles of mass $m$) to cause the mass flux into haloes on larger scales $m_h+m$ and the next merging along the chain. This facilitates a continuous mass cascade from small to large scales. A scale-independent mass flux $\varepsilon_m$ is expected for haloes in the mass propagation range ($<m_h^*$). Mass cascaded from small scales is simply propagated in the propagation range and consumed to grow haloes with mass $>m_h^*$ in the deposition range. } 
\label{fig:3}
\end{figure}

To derive the halo mass function and density profiles, we first introduce the relevant context and background. In CDM cosmology, haloes are continuously merging with small structures (mass accretion). This facilitates an inverse mass cascade in halo mass space, i.e. a continuous mass transfer from small to large mass scales ("inverse") to allow hierarchical structure formation (see Fig. \ref{fig:3}). To explain this, we first identify all haloes in entire system and then group them according to their mass $m_h$. In simulation, a clear definition of halo is required to identify these haloes. This definition is usually related to a critical density $\delta_c$ from a simplified collapse model. At this step, we just treat haloes as existing objects without triggering a specific halo definition. In Fig. \ref{fig:3}, halo of mass $m_h$ merging with a single merger of mass $m$ results in a new halo of mass $m_h+m$. This causes a continuous mass flux from small to large scales along the chain of merging, i.e. an inverse mass cascade at a rate of $\varepsilon_m$.

Next, the mass of entire halo group ($m_g$) including all haloes of the same mass $m_h$ is $m_g=N_h m_h$, where $N_h$ is the number of haloes in that group. Now let's consider the most dominant and frequent merging, i.e. the merging with a single merger (or a single particle of mass $m$) in Fig. \ref{fig:3}, where $\tau_h$ is the average waiting time of a given halo group, i.e. the average time interval between two subsequent merging events involving single mergers with any one halo in the same group. Therefore, the rate of mass transfer (or cascade) from mass scale $m_h$ to scale $m_h+m$ during the time interval $\tau_h$ should be 
\begin{equation}
\varepsilon_m = -\frac{m_h}{\tau_h(m_h,a)}=-\frac{\partial }{\partial t} \left[M_{h} \left(a\right)\int _{m_{h} }^{\infty }f_{M} \left(m,a \right) dm\right], 
\label{eq:4-41}
\end{equation}
i.e. the entire halo mass $m_h$ is transferred to a larger scale in a time interval $\tau_h$. This equals the rate of change for total mass in all haloes greater than $m_h$. Here $M_h(a)$ is the total mass in all haloes, $f_M(m_h,a)=F_M/\rho_0$ (see Eq. \eqref{eq:1-1}) is the probability distribution of total halo mass $M_h$ with respect to $m_h$. The integration gives the total mass in all haloes greater than scale $m_h$. The 'minus' sign stands for the "inverse" cascade from small to large scales.  

When self-gravitating collisionless system reaches a statistically steady state, this rate of mass transfer must be scale independent (i.e. $\varepsilon_m$ is independent of $m_h$).
If this is not the case, there would be a net accumulation of mass at some intermediate mass scale below $m_{h}^{*} $. We exclude this possibility because we require statistical structures of haloes to be self-similar and scale free for haloes smaller than $m_{h}^{*}$. This leads to the rate of mass cascade $\varepsilon_m$ independent of mass scale $m_h$ up to a critical mass $m_h^*$ \citep{Xu:2021-Inverse-mass-cascade-mass-function}. Therefore, taking the derivative of Eq. \eqref{eq:4-41} with respect to $m_h$ leads to
\begin{equation}
\begin{split}
&\frac{\partial \varepsilon_{m}}{\partial m_{h}}=\frac{\partial \left[M_{h} \left(a\right)f_{M} \left(m_{h} ,m_{h}^{*} \right)\right]}{\partial t} =\frac{\partial m_{g} \left(m_{h} ,a\right)}{m_{p} \partial t}=0, \\
&m_g(m_h,t)=M_h(a)f_M(m_h,m_h^*)m_p \equiv m_g(m_h),
\end{split}
\label{eq:4-5}
\end{equation}
where $m_g=N_hm_h$ is the halo group mass, $m_p$ is mass of a single particle (mass resolution in N-body simulation). 

\begin{figure}
\includegraphics*[width=\columnwidth]{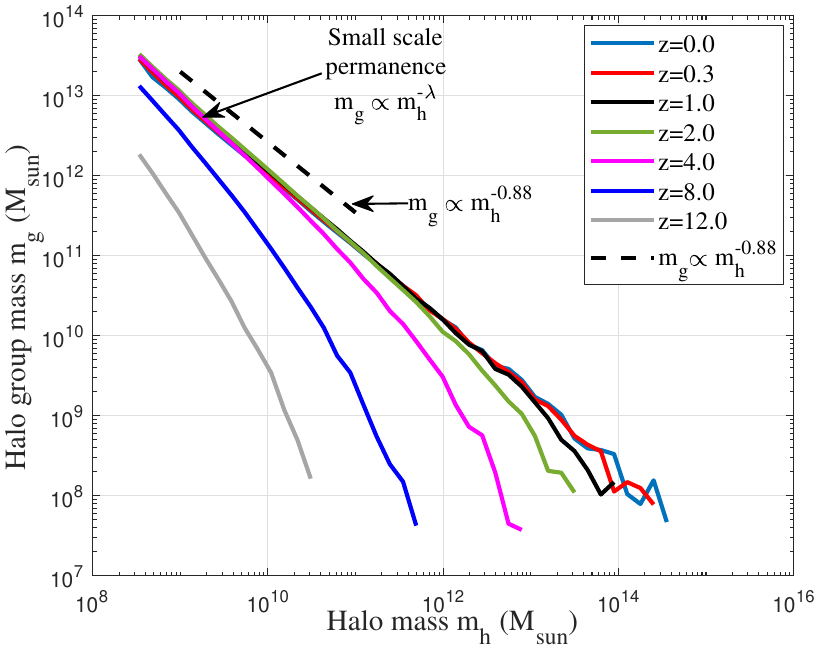}
\caption{The variation of halo group mass $m_g$ with halo mass $m_h$ at different redshift $z$ from Illustris-1-Dark simulation. Figure demonstrates the small scale permanence of group mass $m_g$ in mass space. Once the statistically steady state is established ($z<8$), rate of inverse mass cascade $\varepsilon_m$ becomes scale independent such that the halo group mass $m_g$ at different redshift $z$ collapse to a time independent power-law $m_g\propto m_h^{-\lambda}$ (Eq. \eqref{eq:4-3}) at small mass scale (propagation range) with halo geometry parameter $\lambda\approx 0.88$.}
\label{fig:4-4}
\end{figure}

Here the scale-independent $\varepsilon_m$ requires the halo group mass $m_g(m_h,t)\equiv m_g(m_h)$ to be independent of time, i.e. a "small scale permanence" where the group mass $m_g$ of different halo masses $m_h$ and different redshifts $z$ should collapse on to a common scaling law (Eq. \eqref{eq:4-3} and Fig. \ref{fig:4-4}). Once the statistically steady state is established, the rate of mass cascade $\varepsilon_m$ becomes scale-independent. The halo group mass $m_g$ in propagation range becomes time independent due to scale-independent $\varepsilon_m$. Mass is simply injected at the smallest scale (scale of single mergers), propagated to larger scales in propagation range ($m_h<m_h^*$), and consumed to grow haloes in deposition range ($m_h>m_h^*$). Halo group mass $m_g(m_h)$ is constant in time for haloes $m_h<m_h^*$, and grows with time for haloes $m_h>m_h^*$. Similarly, due to scale-independent energy cascade, the "small scale permanence" for halo density profile will be identified in Section \ref{sec:5} (Fig. \ref{fig:4-3}).

To validate this concept, Fig. \ref{fig:4-4} presents results from large scale cosmological Illustris simulation (Illustris-1-Dark) \citep{NELSON:2015-The-illustris-simulation}. \rev{Illustris is a suite of large volume cosmological DM-only and hydrodynamical simulations. The selected Illustris-1-Dark is the DM-only simulation of 106.5Mpc$^3$ cosmological volume with 1820$^3$ DM particles for the highest resolution. Each DM particle has a mass around $7.6\times 10^6 M_{\odot}$. The gravitational softening length is around 1.4kpc. Haloes in simulation were identified by a standard friends-of-friends (FoF) algorithm with linking length parameter b = 0.2 and halo center placed at the minimum of the gravitational potential of entire halo. Simulation has cosmological parameters of a total matter density $\Omega_m=0.2726$, dark energy density $\Omega_{DE}=0.7274$ at $z=0$, and a dimensionless Hubble constant $h=0.704$.}

Next, if we focus on a given halo in a halo group, the waiting time $\tau_g$ for that particular halo to merge with a single merger should be different and much greater than $\tau_h$ (the waiting time for entire group). Here $\tau_g$ is expected to be inversely proportional to the surface area of that halo. The larger surface area $S_h$, the more likely for that halo to merge with a single merger, and the smaller waiting time $\tau_g$. Therefore, for haloes with a given mass $m_{h}$, this waiting time $\tau_{g} \propto S_h^{-1} \propto {m_{h}^{-\lambda}}$, where $\lambda$ is a key halo geometry parameter. Intuitively, $\lambda \approx 2/3$ for large haloes (i.e. $S_h \propto m_h^{2/3}$). This is also equivalent to the waiting time $\tau_g\propto \Phi^{-1}$, where $\Phi \propto Gm_h/r_h$ is the gravitational potential and $r_h \propto m_h^{1/3}$ is the size of halo. The greater halo gravitational potential $\Phi$, the larger velocity dispersion $\sigma^2$ from virial theorem (or halo temperature), the smaller waiting time $\tau_g$, and the more frequently halo merging with single mergers. Particle waiting time is dependent on its local potential. This will be used for deriving halo density profile in Section \ref{sec:6}.

Depending on the number of haloes $N_h$ in a given halo group, the two waiting times $\tau_g$ and $\tau_h$ are related to each other as 
\begin{equation}
\tau_h=-\frac{m_h}{\varepsilon_m}=\frac{\tau_g}{N_h} \propto N_h^{-1} m_h^{-\lambda} \quad \textrm{and} \quad m_g=N_h m_h\propto m_h^{-\lambda}. 
\label{eq:4-3}
\end{equation}
Again, due to scale-independent rate of mass cascade $\varepsilon_m$ (not varying with $m_h$ in propagation range), Eq. \eqref{eq:4-3} requires the number of haloes $N_h \propto m_h^{-1-\lambda}$ for any given mass $m_h$, or equivalently a power-law group mass $m_g=N_h m_h\propto m_h^{-\lambda}$ at small mass scales, i.e. the small scale permanence in Fig. \ref{fig:4-4}. In the same figure, we obtain $\lambda\approx 0.88$ for Illustris simulation and number of haloes in halo group $N_h\propto m_h^{-1.9}$ that is in good agreement with other work \citep{Bullock:2017-Small-Scale-Challenges-to-the}. 

To summarize, the mass cascade at statistically steady state involves two ranges, the propagation and deposition range. The propagation range for haloes with mass $m_{h}<m_{h}^{*}$ involves a sequence of merging with single mergers (the smallest structure) to simply propagate mass to larger scales. In this range, the rate of mass transfer $\varepsilon_m$ is independent of halo mass $m_h$ and halo group mass $m_g$ is constant in time. The deposition range ($m_{h}>m_{h}^{*}$) involves the consumption (deposition) of mass cascaded from scales below $m_{h}^{*}$ to grow haloes above $m_{h}^{*}$ (Fig. \ref{fig:3}). Therefore, the inverse mass cascade can be described as: "Little halos have big halos, That feed on their mass; And big halos have greater halos, And so on to growth."

In addition, haloes possess finite kinetic and potential energy. Accompanied by the mass cascade, there exists a simultaneous energy cascade across haloes of different masses \citep{Xu:2022-Postulating-dark-matter-partic, Xu:2021-Inverse-and-direct-cascade-of-}. The rate of energy cascade $\varepsilon_u \propto \varepsilon_m \left\langle \sigma^2 \right\rangle /M_h \propto -H\left\langle \sigma^2 \right\rangle$, where $\left\langle \sigma^2 \right\rangle$ is the mean kinetic energy of all particles in all haloes. The specific rate of energy cascade per unit mass ($\varepsilon_u<0$ for inverse energy cascade) can be estimated from the time variation of velocity dispersion $u_0^2$ for all dark matter particles,  
\begin{equation} 
\label{eq:4-4} 
\begin{split}
\varepsilon _{u} = -\frac{3}{2} \frac{u_{0}^{2} }{t_{0} }\approx -4.6\times 10^{-7} \frac{m^{2} }{s^{3}}, 
\end{split}
\end{equation} 
where $u_0 \approx 350km/s$ from N-body simulation and $t_0$ is the current age of universe \citep{Xu:2023-Universal-scaling-laws-and-density-slope}.

Therefore, similar to the mass cascade in propagation range, there exist an inverse (kinetic) energy cascade from small to large scales with a constant rate $\varepsilon_u$. In this range of scales, the small scale structures evolve so fast and do not feel the slowly evolving large scale structures directly except through constant rate $\varepsilon_u$. This description indicates that relevant quantities in this range of scales should be determined by and only by $\varepsilon_{u}$ (${m^{2}/s^{3}}$), gravitational constant $G$ ($m^3/kg\cdot s^2)$, and the relevant length scale \textit{r}. By a simple dimensional analysis, the halo mass enclosed within $r$ and corresponding halo density should follow the scaling \citep{Xu:2023-Universal-scaling-laws-and-density-slope}
\begin{equation}
\label{eq:28} 
m_r(r) \propto \varepsilon _{u}^{2/3}G^{-1}r^{5/3} \quad \textrm{and} \quad \rho_r(r) \propto \varepsilon_{u}^{2/3}G^{-1}r^{-4/3},
\end{equation} 
i.e. the 5/3 law and -4/3 law. These results can be demonstrated and confirmed by both N-body simulations (Figs. \ref{fig:5-1} to \ref{fig:5-4}) and halo density profiles from random walk in Section \ref{sec:6} (Eq. \eqref{eq:21}).

\section{Double-\texorpdfstring{$\lambda$}{} halo mass function}
\label{sec:4}
To derive halo mass function, the inverse mass cascade can be transformed into a halo random walk in mass space that mimics the random work of particles for diffusion problem. Just similar to the particle diffusion, we can derive the relevant Fokker-Planck equation and corresponding solution, from which halo mass function can be analytically solved. This is not just mathematically convenient, but reveals some fundamental aspects of halo mass function as an intrinsic property of self-gravitating collisionless system.

As shown in Fig. \ref{fig:3}, haloes are continuously migrating in mass space from one scale ($m_h$) to neighboring scale ($m_h+m$) by merging with single mergers. This leads to a probability distribution to find a halo at a given mass. The waiting time (or jumping frequency) for a given halo to migrate from a given mass $m_h$ to neighboring mass $m_h+m$ is $\tau_g$ in Eq. \eqref{eq:4-3}. Different from the standard random walk with a constant waiting time, the halo waiting time $\tau_{g}$ is dependent on the mass of halo, i.e. a position-dependent $\tau_g$ (Eq. \eqref{eq:4-3}). For halo with a given mass $m_{h}$, the waiting time $\tau_{g} \propto {m_{h}^{-\lambda}}$, where $\lambda$ is a key halo geometry parameter we discussed. 

First, the random walk of haloes in mass space describes the stochastic variation of the mass of a given halo due to continuous merging with single mergers of mass $m$. Following the Langevin equation, we can write a stochastic equation for halo mass $m_h$ \citep{Xu:2021-Inverse-mass-cascade-mass-function}
\begin{equation}
\label{eq:5} 
\frac{\partial m_{h} \left(t\right)}{\partial t} =\sqrt{2D_{p} \left(m_{h} \right)} \varsigma \left(t\right) \propto \frac{m}{\tau_g},        
\end{equation} 
where $m/\tau_g$ represents the average rate of mass change. For a power-law waiting time $\tau_{g} \propto {m_{h}^{-\lambda}}$, we find the position-dependent diffusivity should take the form of
\begin{equation} 
\label{eq:6} 
D_{p} \left(m_{h} \right)=D_{p0}(t) m_{h}^{2\lambda}.         
\end{equation} 
Here $D_{p0}(t)$ is a proportional constant for diffusivity $D_{p}$. The white Gaussian noise $\varsigma (t)$ satisfies the covariance $\langle \varsigma (t)\varsigma (t^{'})\rangle =\delta (t-t^{'})$ with a zero mean $\langle \varsigma (t)\rangle =0$. Equation \eqref{eq:5} describes the stochastic evolution of halo mass $m_{h} $ with a waiting time $\tau_{g}(m_{h})\propto {m_{h}^{-\lambda}}$. 

Second, in Stratonovich interpretation \citep{Stratonovich:1966-A-new-representation-for-stoch}, the Langevin equation (Eq. \eqref{eq:5}) yields to a distribution function $P_{h} \left(m_{h} ,t\right)$ satisfying the Fokker-Planck equation (resembling particle diffusion)
\begin{equation} 
\label{eq:7} 
\frac{\partial P_{h}(m_h,t) }{\partial t}=D_{p0} \frac{\partial }{\partial m_{h} } \left[m_{h}^{\lambda } \frac{\partial }{\partial m_{h} } \left(m_{h}^{\lambda } P_{h}(m_h,t) \right)\right],
\end{equation} 
which describes the evolution of probability function $P_h$ for halo mass $m_{h}$ in mass space. Obviously, the halo mass function $f_M(m_h,t)$ is exactly the distribution function $P_{h}$, i.e. $f_{M}\equiv P_{h}$. 

Finally, solution to Eq. \eqref{eq:7}, i.e. the halo mass function, is a stretched Gaussian with an exponential cut-off for large $m_{h}$ and a power-law behavior for small $m_{h}$,
\begin{equation}
\label{eq:8} 
f_{M} \left(m_{h},t\right)=\frac{m_{h}^{-\lambda} }{\sqrt{\pi D_{p0} t} } \exp \left[-\frac{m_{h}^{2-2\lambda } }{4\left(1-\lambda \right)^{2} D_{p0} t} \right].      
\end{equation} 
The mean square displacement in mass space is 
\begin{equation} 
\label{eq:9} 
\begin{split}
&\left\langle m_{h}^{2} \right\rangle =\int _{0}^{\infty }f_{M} \left(m_{h} ,t\right) m_{h}^{2} dm_{h}\\
&=\frac{1}{\sqrt{\pi } } \Gamma \left(\frac{3-\lambda }{2-2\lambda } \right)\left[4\left(1-\lambda \right)^2 D_{p0} t\right]^{\frac{1}{1-\lambda } } \equiv \gamma _{0} m_{h}^{*2}.
\end{split}
\end{equation} 
where $m_{h}^{*}(t)$ is the critical mass scale and $\gamma _{0} $ is just a proportional constant. With the exponent of ${1/\left(1-\lambda \right)} \ge 1$ in Eq. \eqref{eq:9}, it is clear that the random walk of haloes in mass space is of a super-diffusion nature. Now $f_{M} \left(m_{h} ,t\right)$ (Eq. \eqref{eq:8}) can be rewritten in terms of $m_{h}^{*}$
\begin{equation}
\label{eq:10} 
f_{M} \left(m_{h} ,t\right)=\frac{\left(1-\lambda \right)}{m_h^*\sqrt{\pi \eta _{0} } } \left(\frac{m_{h}^{*} }{m_{h}} \right)^{\lambda } \exp \left[-\frac{1}{4\eta _{0} } \left(\frac{m_{h} }{m_{h}^{*} } \right)^{2-2\lambda } \right],     
\end{equation} 
where the dimensionless constant
\begin{equation} 
\label{eq:11} 
\eta _{0} =\frac{1}{4} \left[\frac{\gamma _{0} \sqrt{\pi } }{\Gamma \left({\left(3-\lambda \right)/\left(2-2\lambda \right)} \right)} \right]^{1-\lambda } .        
\end{equation} 

The time dependence of $f_M$ is absorbed into $m_h^*$. 
Intuitively, $\lambda \approx 2/3$ for large haloes in deposition range with low concentration, whose central structures are still dynamically adjusted due to fast mass accretion. While for small haloes with high concentration (propagation range), the mass accretion is slow and inner structure is stable \citep{Zhao:2009-Accurate-Universal-Models-for-}. These small haloes can be treated as fractal objects with a fractal surface dimension $D_h \le 3$. The geometry parameter $\lambda=D_h/3$ can be greater than $2/3$ (see Fig. \ref{fig:4-4}). These high concentration low mass haloes are usually found in denser environments \citep{Maccio:2007-Concentration--spin-and-shape-}. The denser environment might lead to a rougher halo surface and higher surface fractal dimension $D_h$. Therefore, two different $\lambda$ (i.e. double-$\lambda$) are required for two ranges (propagation range with $m_h<m_h^*$ and deposition range with $m_h>m_h^*$) due to different halo properties and surrounding environments. The single-$\lambda$ halo mass function in Eq. \eqref{eq:10} can be naturally generalized to a double-$\lambda$ halo mass function with $\lambda_{1}$ and $\lambda_{2}$ for propagation and deposition ranges, respectively. Therefore, the double-$\lambda$ mass function reads
\begin{equation} 
\label{eq:12} 
\begin{split}
f_{M} (m_{h},a)&=\left(2\sqrt{\eta _{0} } \right)^{-q} \frac{2\left(1-\lambda _{1} \right)}{q\Gamma \left({q/2} \right)}\\ 
&\cdot \left(\frac{m_h^*}{m_h}\right)^{\lambda _{1}}\frac{1}{m_{h}^{*} } \exp \left[-\frac{1}{4\eta _{0} } \left(\frac{m_{h}}{m_{h}^{*}} \right)^{2-2\lambda _{2} } \right]. 
\end{split}
\end{equation} 
By introducing variable $\nu =(m_h/m_h^*)^{2/3}$, the three parameter double-$\lambda$ mass function can be finally written as,
\begin{equation} 
\label{eq:13} 
f_{D\lambda} (\nu)=\frac{p\left(2\sqrt{\eta _{0} }\right)^{-q} }{\Gamma \left({q/2} \right)}{\nu}^{\frac{pq}{2}-1} \exp \left(-\frac{{\nu}^p }{4\eta _{0} } \right),       
\end{equation} 
where model parameters $p$ and $q$ have clear physical meaning. Both are related to halo geometry parameters $\lambda_1$ and $\lambda_2$ as,  
\begin{equation} 
\label{eq:14} 
p=3\left(1-\lambda_2\right) \quad \textrm{and} \quad  q=\frac{\left(1-\lambda _{1} \right)}{\left(1-\lambda _{2} \right)}.
\end{equation} 

Clearly, Eq. \eqref{eq:13} reduces to the Press-Schechter (PS) mass function if $\lambda_1=\lambda_2=2/3$ and $\eta _{0}={1/2}$. However, the derivation of double-$\lambda$ mass function does not rely on any collapse model (spherical or ellipsoidal). The critical overdensity $\delta_{c}$ from collapse model is not required in this formulation. In simulation, haloes are usually defined using the critical overdensity $\delta_c$ to compute the halo mass function. The derivation of double-$\lambda$ mass function of Eq. \eqref{eq:13} does not depend on the exact definition of halo. Different definitions of halo in simulation might affect both halo mass $m_h$ and the critical mass $m_h^*$, but not the ratio $\nu =(m_h/m_h^*)^{2/3}$, and therefore not the double-$\lambda$ halo mass function. More importantly, $\lambda_1=\lambda_2=2/3$ or $p=q=1$ is a natural result of current theory. This formulation reveals that the halo mass function in the form of Eq. \eqref{eq:13} is an intrinsic property of self-gravitating collisionless dark matter system that is independent of spherical or ellipsoidal collapse models.

\begin{figure}
\includegraphics*[width=\columnwidth]{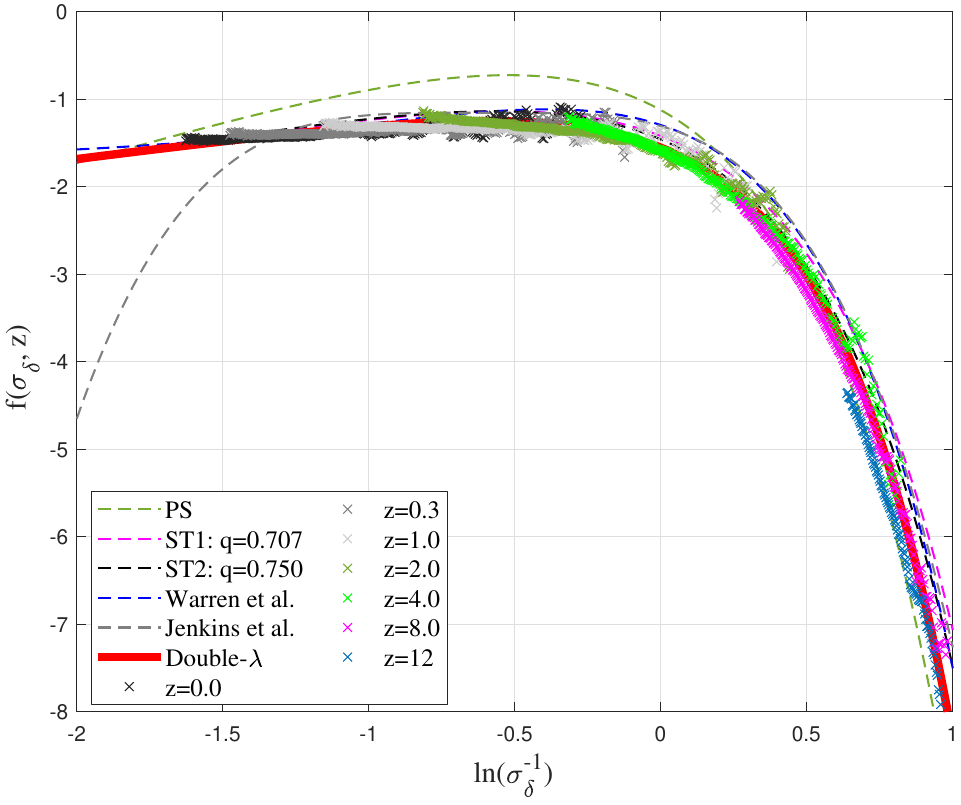}
\caption{Comparison between different halo mass functions $\textbf{f}(\sigma_{\delta}, z)$ and simulation at different redshift \textit{z}. The PS mass function overestimate the mass in small haloes and underestimates the mass in large haloes. The fitted JK mass function matches simulation only in a given range with large deviation for small mass haloes. The WR mass function deviates at small mass with a limit $\textbf{f}(\sigma_{\delta}^{-1}\rightarrow0,z)=-1.695$. The double-$\lambda$ mass function (Eq. \eqref{eq:13}) with best fitting parameters $\eta_{0}=1.162$, $q=0.365$, and $p=1.185$ (or $\lambda_1=0.856$ and $\lambda_2=0.605$) matches the simulation and is slightly better than ST mass functions at large halo mass.}
\label{fig:3-2}
\end{figure}

The halo geometry exponent $\lambda$ has a fundamental meaning to relate halo surface area (or effective mass accretion area) to its mass. The cosmology and redshift dependence of $\lambda_1$ and $\lambda_2$ can be systematically studied by fitting the model to the simulation data of different cosmologies, similar to the study in \citet{Bocquet:2020-The-Mira-Titan-Universe} and \citet{Castro:2022-Calibration-of-the-halo-mass-function}. 

Alternatively, similar to the scale radius $r_s$ for halo density where logarithmic density slope is -2, we may introduce a scale mass $m_{hs}$ where logarithmic slope $\partial\ln(f_M)/\partial\ln(m_h) = -1$ such that $m_{hs} = (2\eta_0 q)^{3/(2p)}m_h^*$ from Eq. \eqref{eq:12}. With a new scaled variable $\bar\nu=(m_h/m_{hs})^{2/3}$, mass function in Eq. \eqref{eq:13} can be further simplified with $p$ and $q$ as the only two parameters
\begin{equation} 
\label{eq:13-2} 
f_{D\lambda} (\bar\nu)=\frac{p(q/2)^{q/2}}{\Gamma \left({q/2} \right)}{\bar\nu}^{\frac{pq}{2}-1} \exp \left(-\frac{q}{2}{\bar\nu}^p  \right).  
\end{equation} 

To validate the derived double-$\lambda$ mass function, we presents results from Illustris simulation (Illustris-1-Dark) \citep{NELSON:2015-The-illustris-simulation}. Figure \ref{fig:3-2} presents the halo mass function $\textbf{f}(\sigma_{\delta}, z)$ in Eq. \eqref{eq:1-1}. The best fit of double-$\lambda$ mass function to the simulation data at all z gives values of $\eta_{0}=1.162$, $q=0.365$, and $p=1.185$ (Fig. \ref{fig:3-2}), which leads to $\lambda_1=0.856$ and $\lambda_2=0.605$ from Eq. \eqref{eq:14} for the propagation and deposition ranges, respectively. This leads to a slope of $-\lambda_1-1\approx-1.9$ for halo number density $n(m_h,z) \propto m_h^{-1.9}$ (Eq. \eqref{eq:4-3}), in very good agreement with Fig. \ref{fig:4-4} and other work \citep{Bullock:2017-Small-Scale-Challenges-to-the}. Compared to predicted value of $\lambda=2/3$ for matter dominant universe, the effect of dark energy in Illustris simulations seems to enhance the value of $\lambda_1$ and decrease the value of $\lambda_2$, reflecting the changes in environments and halo properties due to the presence of dark energy and accelerated expansion. 

\begin{figure}
\includegraphics*[width=\columnwidth]{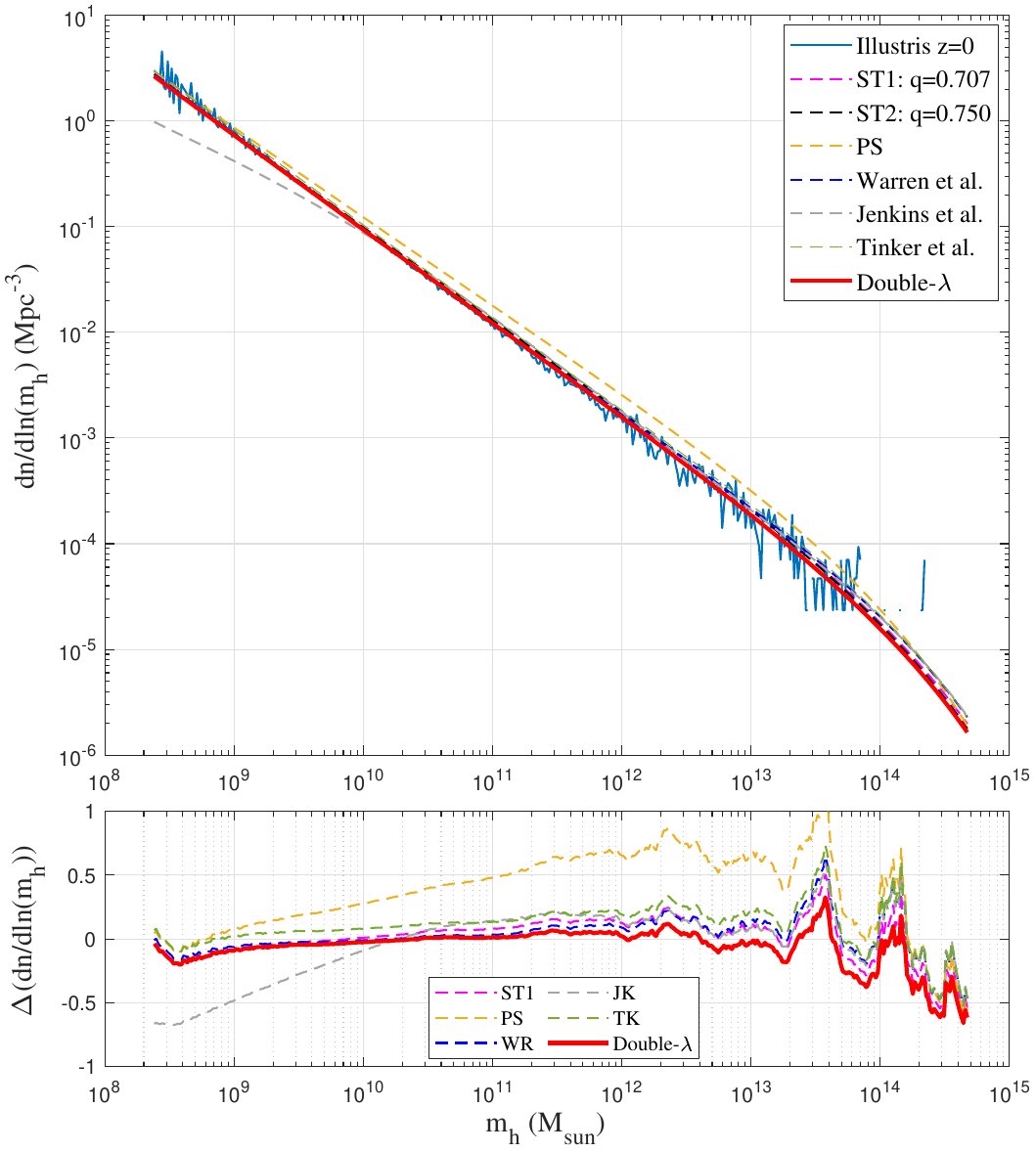}
\caption{Comparison of mass functions with Illustris-1-Dark simulation (solid blue) at \textit{z}=0. The PS mass function overestimates mass in small haloes. The fitted JK mass function matches simulation only in a given range. The double-$\lambda$ mass function (Eq. \eqref{eq:13}) matches both simulation and the ST and WR mass functions at \textit{z}=0. Bottom plot presents the relative errors between simulation and different mass functions.}
\label{fig:2-1}
\end{figure}

The PS mass function overestimate the mass in small haloes and underestimates the mass in large haloes. The JK mass function matches simulation for large mass haloes with large deviation for small haloes. The fitted WR mass function does not satisfy the normalization condition, where $\int _{0}^{\infty }f_{WR}(\nu) d\nu$ diverges. The WR mass function also deviates at small mass with a finite limit $\textbf{f}(\sigma_{\delta}^{-1},z)=-1.695$ for $\sigma_{\delta}\rightarrow \infty$. The ST functions matches the simulation better with $\textbf{f}(\sigma_{\delta},z)\rightarrow \sigma_{\delta}^{2p-1}\approx \sigma_{\delta}^{-0.4}$ for large $\sigma_{\delta}$. For large halo or high redshift, ST mass functions tend to overestimate when compared with simulation, which is also found in other studies \citep{Reed:2003-Evolution-of-the-mass-function,Lukić:2007-The-Halo-Mass-Function}. The double-$\lambda$ mass function is better than ST function for large haloes with $\textbf{f}(\sigma_{\delta},z)\rightarrow \sigma_{\delta}^{-pq}\approx \sigma_{\delta}^{-0.43}$ for $\sigma_{\delta}\rightarrow \infty$.

\begin{figure}
\includegraphics*[width=\columnwidth]{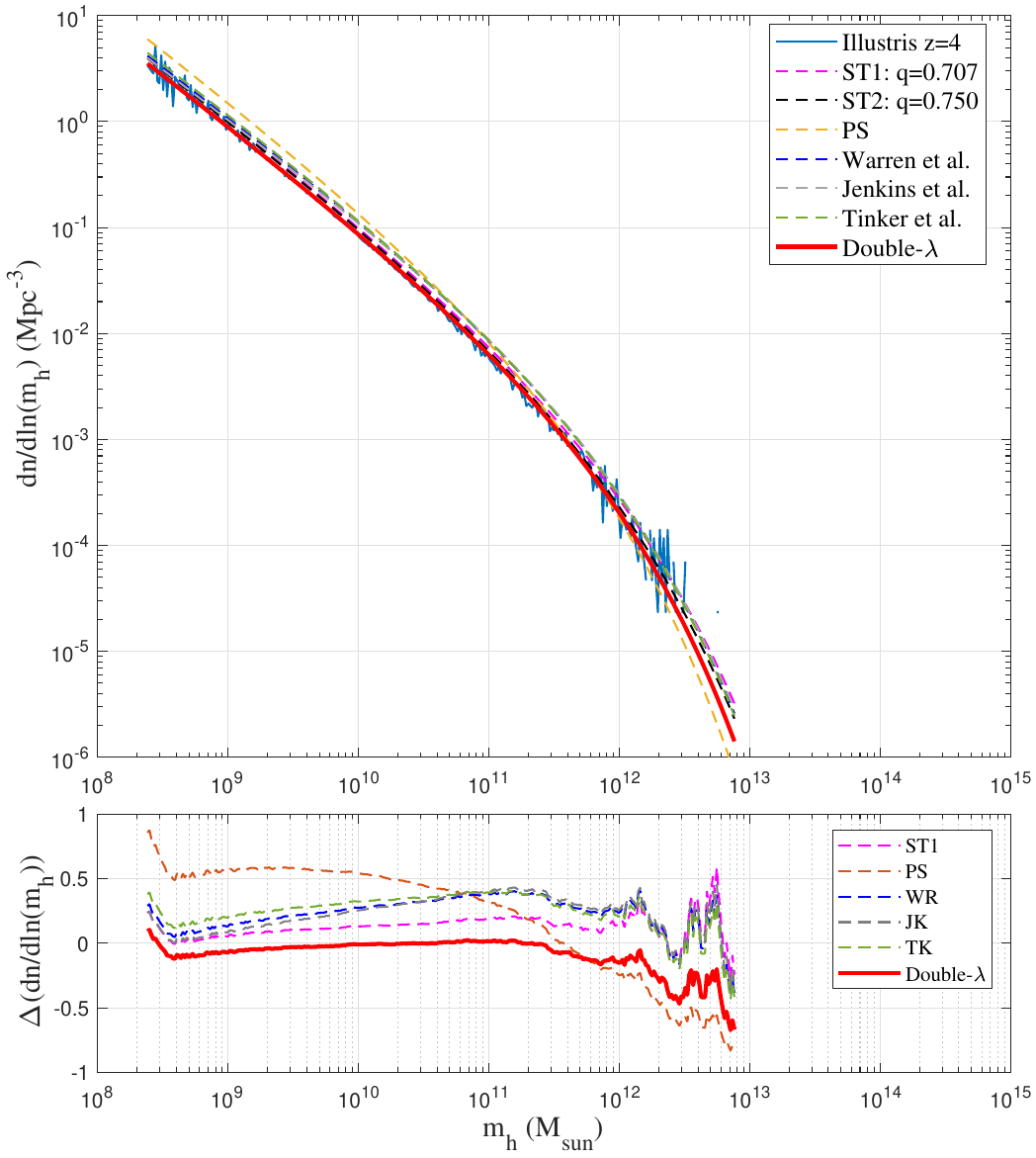}
\caption{Comparison of mass functions with Illustris-1-Dark (solid blue) at \textit{z}=4. The simulation results agree with all mass functions except PS. Double-$\lambda$ mass function (Eq. \eqref{eq:13}) predicts a slightly lower mass in larger haloes.}
\label{fig:2-2}
\end{figure}

\begin{figure}
\includegraphics*[width=\columnwidth]{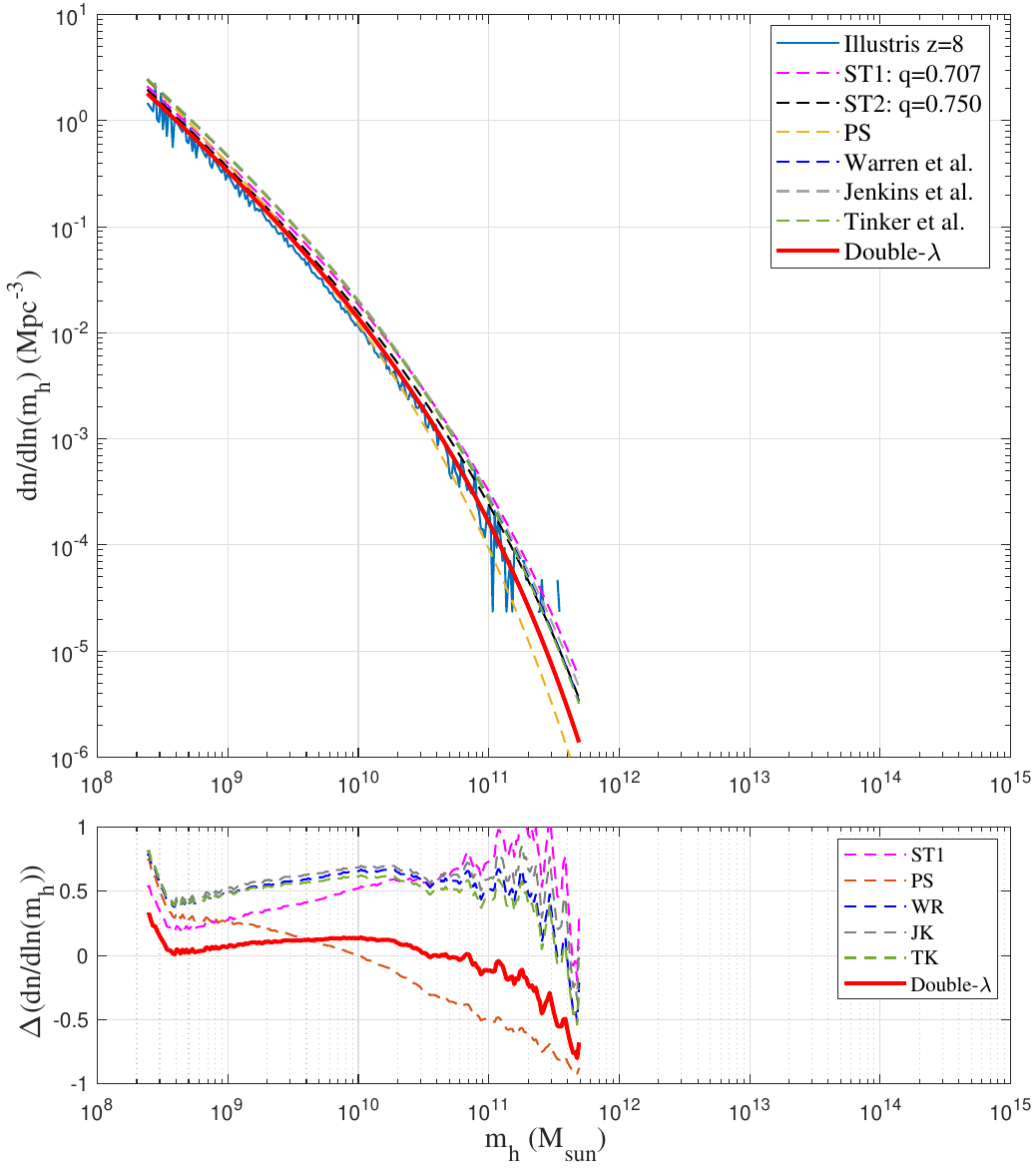}
\caption{Comparison of mass functions with Illustris-1-Dark simulation (solid blue) at \textit{z}=8. The double-$\lambda$ mass function (Eq. \eqref{eq:13}) predicts less mass in larger haloes and slightly better agrees with the simulation.}
\label{fig:2-3}
\end{figure}

\begin{figure}
\includegraphics*[width=\columnwidth]{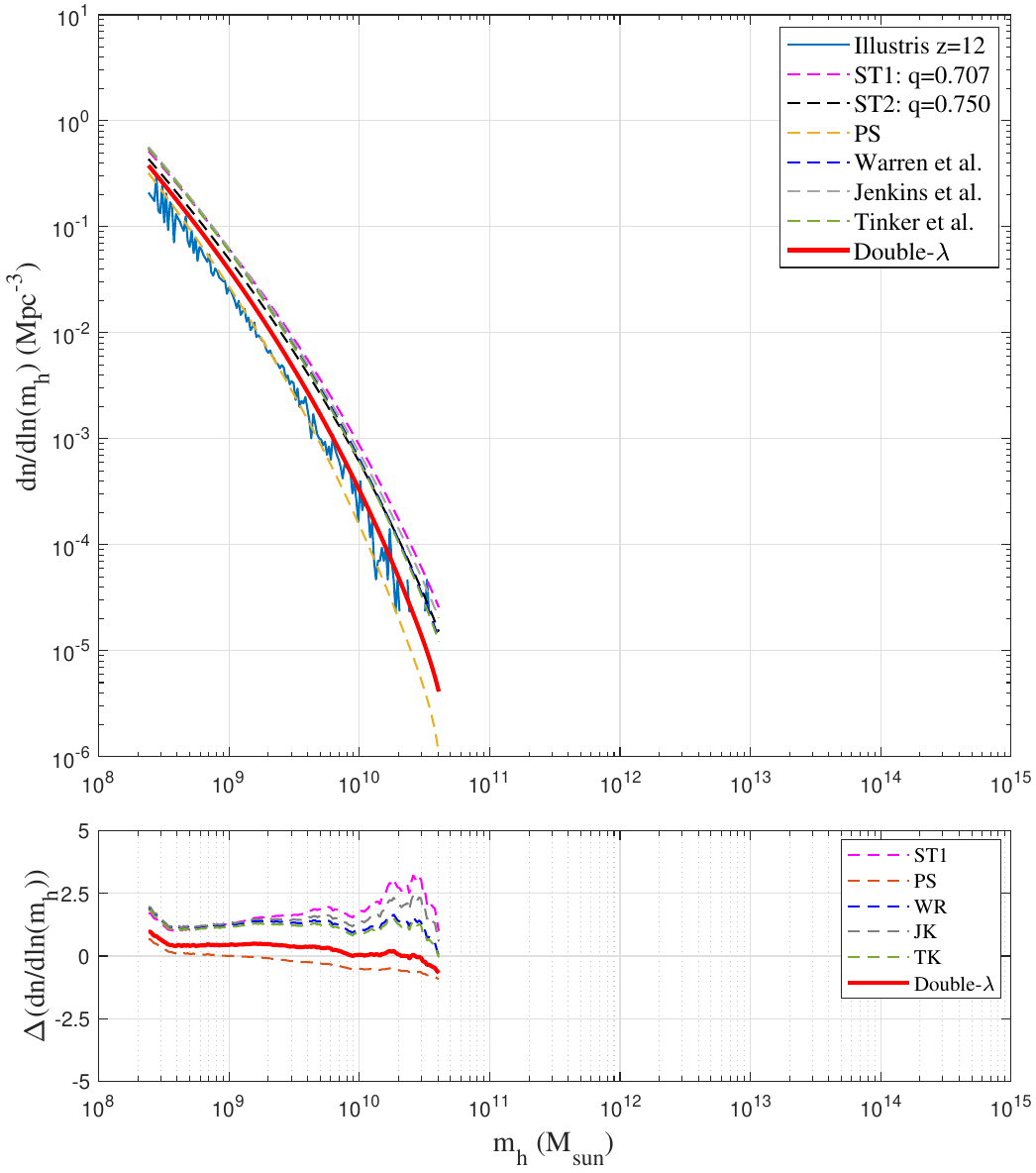}
\caption{Comparison of mass functions with Illustris-1-Dark simulation (solid blue) at \textit{z}=12. Compared to other mass functions, the double-$\lambda$ mass function (Eq. \eqref{eq:13}) predicts less mass in larger haloes and slightly better agrees with the simulation.}
\label{fig:2-4}
\end{figure}

Figures \ref{fig:2-1} to \ref{fig:2-4} present the comparison of halo mass functions $F_M$ in Eq. \eqref{eq:1-1} with simulation results at $z=0, 4, 8, \textrm{and} 12$, as a function of halo mass $m_h$. Relative errors of different mass functions when compared to binned simulation data are also presented in the bottom plots. Similar conclusions can be obtained from these plots, where WR, ST, TK and double-$\lambda$ mass functions agree with simulation at lower redshift. Double-$\lambda$ mass function is slightly better at higher redshifts $z=8$ and 12. 

\section{Mass scale \texorpdfstring{$\lowercase {m_h}^*$}{} and small scale permanence}
\label{sec:5}
The inverse mass cascade and halo mass function (Eq. \eqref{eq:12}) require a critical halo mass scale $m_h^*$ that can be related to halo velocity dispersions from virial theorem
\begin{equation} 
\label{eq:13-3} 
\nu = \left(\frac{m_h}{m_h^*}\right)^{2/3} = \frac{\langle \sigma_v^2(m_h) \rangle}{\langle \sigma_v^2(m_h^*) \rangle} = \frac{\langle \sigma_v^2(m_h) \rangle}{ \sigma_h^2(m_h^*)},
\end{equation}
where $\sigma_v^2(m_h)$ is the velocity dispersion of all DM particles in a halo with a given mass $m_h$, which represents the temperature of that halo. Here $\langle \rangle$ represents the average for all haloes in the same group with same mass $m_h$. In addition, $\sigma_h^2=VAR(V_h)$ is the dispersion (variance) of halo velocity $V_h$ (the mean velocity of all particles in the same halo) for all haloes in the same group, where $\sigma_h^2$ represents the temperature of halo group that is relatively independent of halo mass $m_h$ \citep{Xu:2023-Maximum-entropy-distributions-of-dark-matter,Xu:2021-Inverse-and-direct-cascade-of-}. 

\begin{figure}
\includegraphics*[width=\columnwidth]{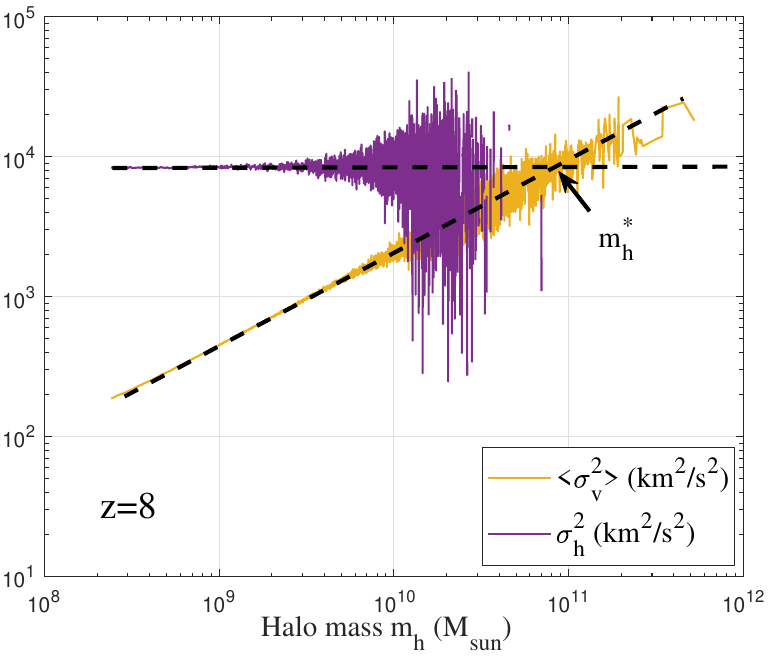}
\caption{The halo velocity dispersions $\langle \sigma_v^2\rangle(m_h) $ and $\sigma_h^2$ at $z=8$ from Illustris-1-Dark simulation. The two velocity dispersions represent the temperature of haloes and temperature of halo groups \citep{Xu:2021-Inverse-and-direct-cascade-of-}. The large fluctuation at large mass scale is due to fewer massive haloes. Here $\langle \sigma_v^2 \rangle \propto m_h^{2/3}$ while $\sigma_h^2$ is relatively independent of $m_h$. The critical halo mass $m_h^*(z=8)=9\times 10^{10}M_{\odot}$ is found by setting $\langle \sigma_v^2 \rangle (m_h^*) = \sigma_h^2$ (Eq. \eqref{eq:13-3}).}
\label{fig:4}
\end{figure}

\begin{figure}
\includegraphics*[width=\columnwidth]{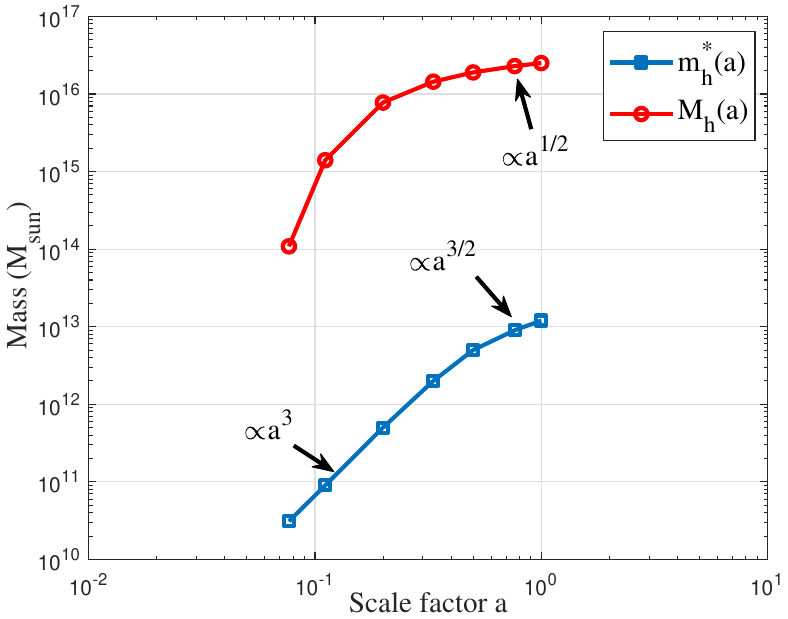}
\caption{The variation of critical halo mass $m_h^*$ and total mass $M_h$ in all haloes with scale factor $a$. Two regimes can be identified. In the linear regime $m_h^*\propto a^3$. In nonlinear regime $m_h^*\propto a^{3/2}$ and $M_h\propto a^{1/2}$, where statistically steady state is established with a scale-independent rate of cascade. Density profiles of haloes with critical mass $m_h^*$ are presented in Fig. \ref{fig:4-3}.}
\label{fig:4-2}
\end{figure}

Figure \ref{fig:4} presents an example of the variation of $\langle\sigma_v^2\rangle$ and $\sigma_h^2$ with $m_h$ at $z=8$, where the critical mass $m_h^*(z=8)=9\times 10^{10}M_{\odot}$ can be determined by setting $\langle \sigma_v^2(m_h^*) \rangle = \sigma_h^2$ in Eq. \eqref{eq:13-3}. We can similarly compute the critical mass $m_h^*$ for other redshifts. The variation of $m_h^*$ with the scale factor $a$ is presented in Fig. \ref{fig:4-2}. In linear regime, $m_h^* \propto a^3$ is expected, while in nonlinear regime $m_h^* \propto a^{3/2}$ \citep{Xu:2021-Inverse-mass-cascade-mass-function}. 

With halo mass function in Eq. \eqref{eq:10} and the small scale permanence for $m_g$ in Eqs. \eqref{eq:4-5}, \eqref{eq:4-3}, and Fig. \ref{fig:4-4}, the halo group mass $m_g=m_hm_p$ ($m_p$ is particle mass) should satisfy 
\begin{equation} 
\label{eq:13-4} 
m_g(m_h,t) = M_h(t)f_Mm_p\propto M_h m_h^*{^{\lambda-1}}m_h^{-\lambda}m_p\equiv m_g(m_h),
\end{equation}
such that the total mass in all haloes $M_h(a)\propto m_h^*{^{1-\lambda}}$ when statistically steady state is established in the nonlinear regime. With $\lambda=2/3$ for $m_h=m_h^*$, $M_h(a)\propto a^{1/2}$ is expected. The time variation of total halo mass $M_h$ is also presented in Fig. \ref{fig:4-2}. 

\begin{figure}
\includegraphics*[width=\columnwidth]{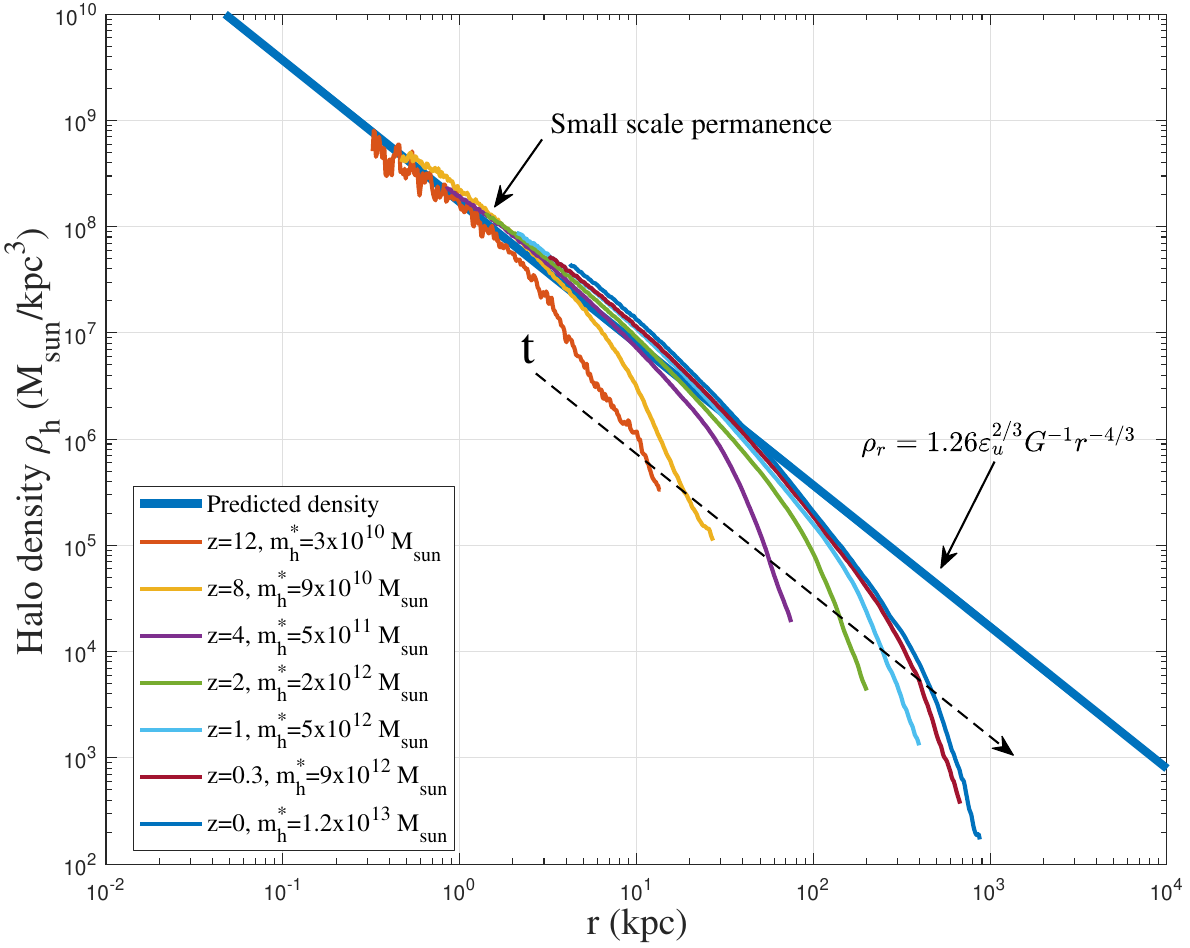}
\caption{The evolution of halo density profiles for haloes with critical mass $m_h^*(z)$. Figure demonstrates the small scale permanence, i.e. the density profiles for haloes with mass $m_h^*$ at different redshifts $z$ collapse at small scale $r$ onto the predicted density scaling (-4/3 law with $\rho_h\propto r^{-4/3}$) from the theory of energy cascade (solid blue line from Eq. \eqref{eq:28}).}
\label{fig:4-3}
\end{figure}

Next, similar to the small scale permanence for group mass $m_g$ in Fig. \ref{fig:4-4}, we will present the small scale permanence for halo density profile. From the scaling laws due to energy cascade, the density scaling $\rho_r \propto r^{-4/3}$ is proposed in Eq. \eqref{eq:28}, which already hints the small scale permanence. To demonstrate this concept, the density profiles for haloes with a critical mass $m_h^*$ at different redshifts are studied first. In Illustris-1-Dark simulation, all haloes with mass between $10^{\pm\Delta}m_h^*$ are identified at different redshifts $z$ with $\Delta=0.1$. The spherical averaged density profile is computed for every halo. The density profile for haloes with critical mass $m_h^*$ is computed as the average density profile for all haloes with mass between $10^{\pm\Delta}m_h^*$. Figure \ref{fig:4-3} presents the time evolution of halo density profiles for haloes with critical mass $m_h^*(z)$. The small scale permanence from energy cascade can be clearly demonstrated as the density profiles for haloes with critical mass at different redshifts all collapse onto the predicted density scaling (blue solid line $\rho_h\propto r^{-4/3}$) on small scales. Finally, if gravity is the only interaction and dark matter is fully collisionless and cold, extending the established scaling in Fig. \ref{fig:4-3} to the smallest length scale and and the earliest time (or the highest $z$) might be able to identify dark matter particle mass, size, lifetime, and many other properties \citep{Xu:2022-Postulating-dark-matter-partic}. 

\section{Double-\texorpdfstring{$\gamma$}{} halo density profile}
\label{sec:6}
The halo density profile can be analytically derived based on a similar idea as deriving halo mass function. Within CDM paradigm, the formation of structures starts from the gravitational collapse of small scale density fluctuations and proceeds hierarchically such that small structures coalesce into large structures in a "bottom-up" fashion. The halo structure is formed hierarchically through a series merging with smaller structures (dominantly with single mergers in Fig. \ref{fig:3}). 

\begin{figure}
\includegraphics*[width=\columnwidth]{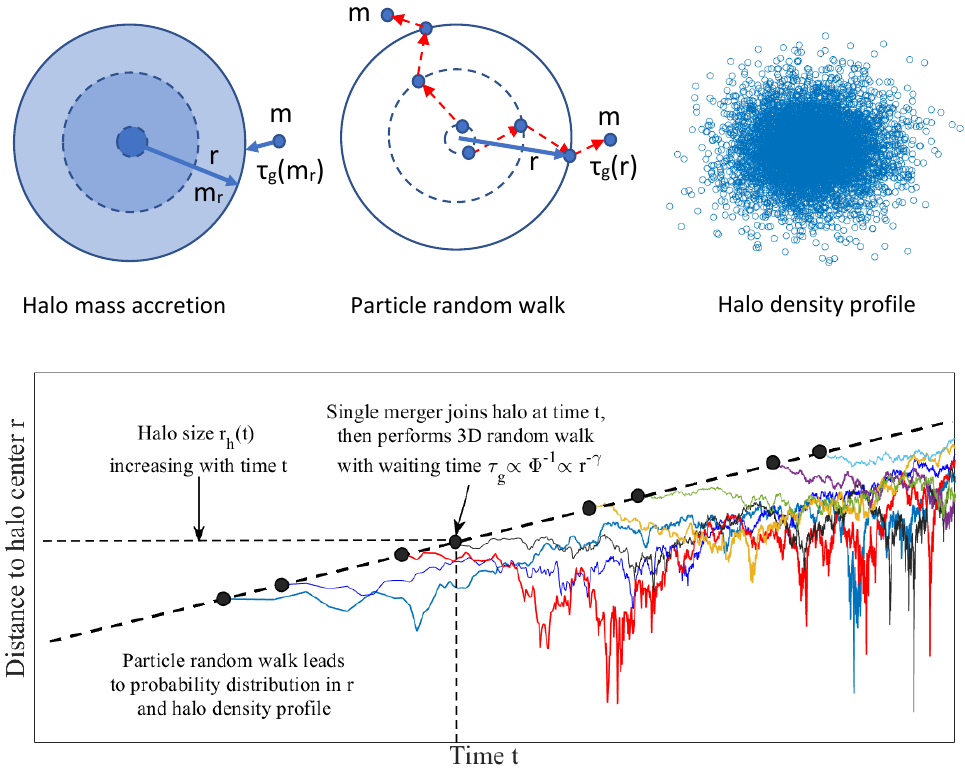}
\caption{Schematic plot of the growth of a given halo in both mass $m_r$ and size $r$ via continuous merging with single mergers, where the waiting time $\tau_g(m_r)\propto m_r^{-\lambda}$. Every merging event corresponds to a single move of particle $m$ in a random walk process, where the waiting time $\tau_g(r)\propto r^{-\gamma}$. Single mergers continuously join halo and perform 3D random walk. Particle distribution from 3D random walk gives rise to the halo density.}
\label{fig:S2}
\end{figure}

Now let us follow the mass accretion history of a given halo in Fig. \ref{fig:S2}, where halo mass $m_r\equiv m_r(t)$ (or halo size $r\equiv r(t)$, the radius enclosing mass $m_r$) continuously varies with time from 0 to $m_r$ (or size from 0 to $r$ ). The mean waiting time of every merging with a single merger $m$ has a simple scaling as $\tau_g \propto m_r^{-\lambda}\propto \Phi^{-1}$, where $\lambda$ is a halo geometry parameter (see Eq. \eqref{eq:4-3}) and $\Phi(r) \propto Gm_r/r$ is the gravitational potential at $r$. In 3D space, halo size $r$ can be related to the position $\boldsymbol{X}_t$ of merger $m$ as $r=\sqrt {\boldsymbol{X}_t \cdot \boldsymbol{X}_t}$. Since both halo mass $m_r(t)$ and $\Phi(r)$ can be related to size $r(t)$, the waiting time $\tau_g$ should also be a function of $r(t)$, which means a varying waiting time dependent on the particle distance $r$ to halo center
\begin{equation} 
\label{eq:18} 
\tau_g(r) \propto \Phi(r)^{-1} \propto r(t)^{-\gamma},
\end{equation} 
where $\gamma$ is an exponent for $r$-dependence of waiting time $\tau_g$, which can be related to the slope of density profile (see Eq. \eqref{eq:21}). 

Since haloes are formed by sequential merging, every DM particle in any halo was a single merger at the time they joined that halo. That particle starts to continuously perform a 3D random walk with a position-dependent waiting time $\tau_g$ dependent on its local potential $\Phi$ or $r$ (Eq. \eqref{eq:18}) right after the merging, where $\Phi(r)$ is determined by the total enclosed mass within $r$. In this regard, halo random walk in mass space is consistent with the particle random walk in 3D space. The random walk of DM particles has a position dependent waiting time $\tau_g \propto \Phi(r)^{-1} \propto r^{-\gamma}$, where $r=\sqrt {\boldsymbol{X}_t \cdot \boldsymbol{X}_t}$ is the distance to halo center. The waiting time is also dependent on the local potential $\Phi(r)$, or from virial theorem, the velocity dispersion $\sigma^2$ that represents the local temperature. Since energy cascade theory predicts the 5/3 law for mass scaling $m_r \propto r^{5/3}$ for the inner region of virialized haloes (see Eq. \eqref{eq:28}), we have potential $\Phi(r) \propto Gm_r/r \propto r^{2/3}$ such that $\gamma=2/3$ from Eq. \eqref{eq:18}. A position dependent waiting time $\tau_g(r)$ is an important feature for hierarchical formation of halo structure. A longer waiting time $\tau_g(r)$ at small $r$ means a more stable core region than the outer region. 

Finally, the particle distribution resulting from this position-dependent random walk in 3D space gives rise to the halo density, as shown in Fig. \ref{fig:S2}. Therefore, to find the halo density profile, we need to derive the particle distribution function due to the random walk in 3D space with $\tau_g(r)\propto r^{-\gamma}$. The 3D particle random walk can be described by a Langevin equation for particle position $\boldsymbol{X}_{t}$ (similar to Eq. \eqref{eq:5} for halo random walk in mass space), 
\begin{equation} 
\label{eq:17} 
\frac{d\boldsymbol{X}_{t} }{dt} = \sqrt{2 D_P(\boldsymbol{X}_t)} \boldsymbol{\xi}\left(t\right).       
\end{equation} 
Due to position-dependent waiting time $\tau_g(r)$, the position-dependent diffusivity reads 
\begin{equation} 
\label{eq:19} 
D_{P}(\boldsymbol{X}_t)=D_{0}(t) r^{2\gamma},         
\end{equation} 
where $D_{0}(t)$ is a proportional constant. The smaller $r$, the smaller diffusivity or longer waiting time, and the higher particle density. In It$\hat{\textrm{o}}$ convention, the 3D Fokker-Planck equation in Cartesian coordinate can be directly obtained for particle distribution function $P_{r} \left(\boldsymbol{X},t\right)$ ($i=1,2,3$ for Cartesian coordinates),
\begin{equation} 
\label{eq:20} 
\begin{split}
\frac{\partial P_{r} \left(\boldsymbol{X},t\right)}{\partial t} = D_0\frac{\partial }{\partial X_i}\left[ \frac{\partial }{\partial X_i}\left(r^{2\gamma}P_{r} \left(\boldsymbol{X},t\right)\right) \right]. 
\end{split}
\end{equation} 
The corresponding solution of Eq. \eqref{eq:20} in spherical coordinate is 
\begin{equation}
\label{eq:21} 
P_{r} \left(r,t\right) = \frac{(2-2\gamma)^{\frac{\gamma-2}{1-\gamma}} r^{-2\gamma}}{4\pi\left(D_0t\right)^{\frac{3-2\gamma}{2-2\gamma}}\Gamma\left(\frac{3-2\gamma}{2-2\gamma}\right)}\exp \left(-\frac{r^{2-2\gamma}}{4(1-\gamma)^2D_0t}\right).        
\end{equation} 
Since the distribution function $P_r(r,t)$ is equivalent to halo density, we find that the parameter $\gamma$ is half of the  density slope at small $r$. 

\begin{figure}
\includegraphics*[width=\columnwidth]{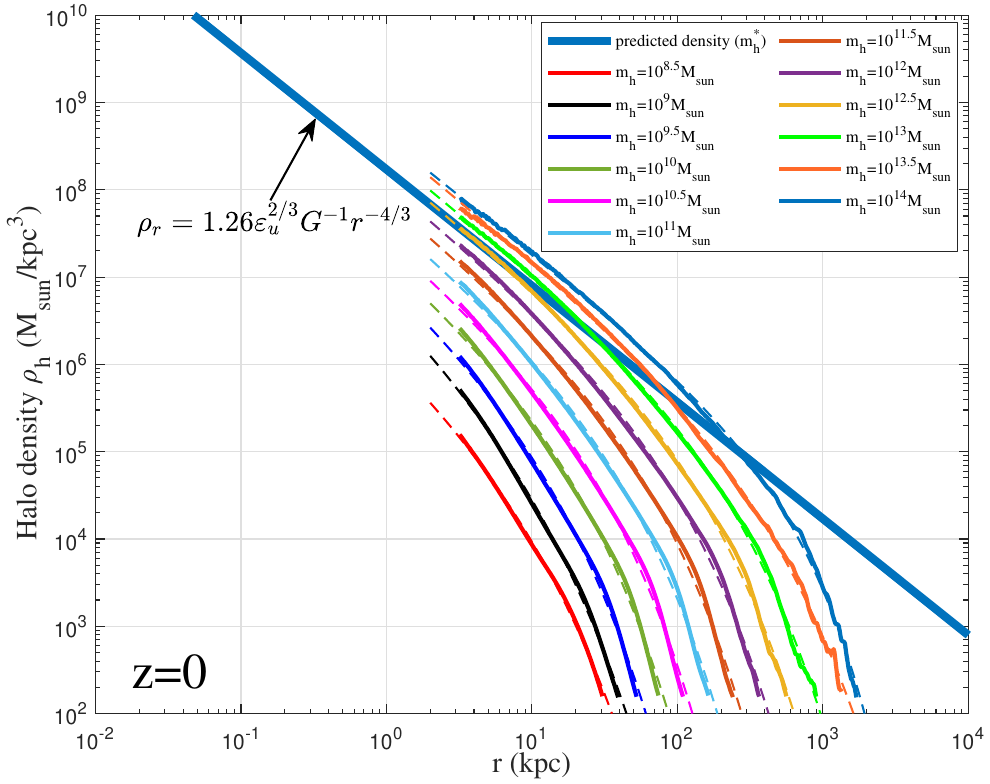}
\caption{Halo density profiles for different halo mass $m_h$ at $z=0$ (solid lines). The predicted scaling law (Eq. \eqref{eq:28}) for halo density is presented as the solid blue line. The double-$\gamma$ density model (Eq. \eqref{eq:29}) was also plotted for all haloes as dashed lines. }
\label{fig:5-1}
\end{figure}

\begin{figure}
\includegraphics*[width=\columnwidth]{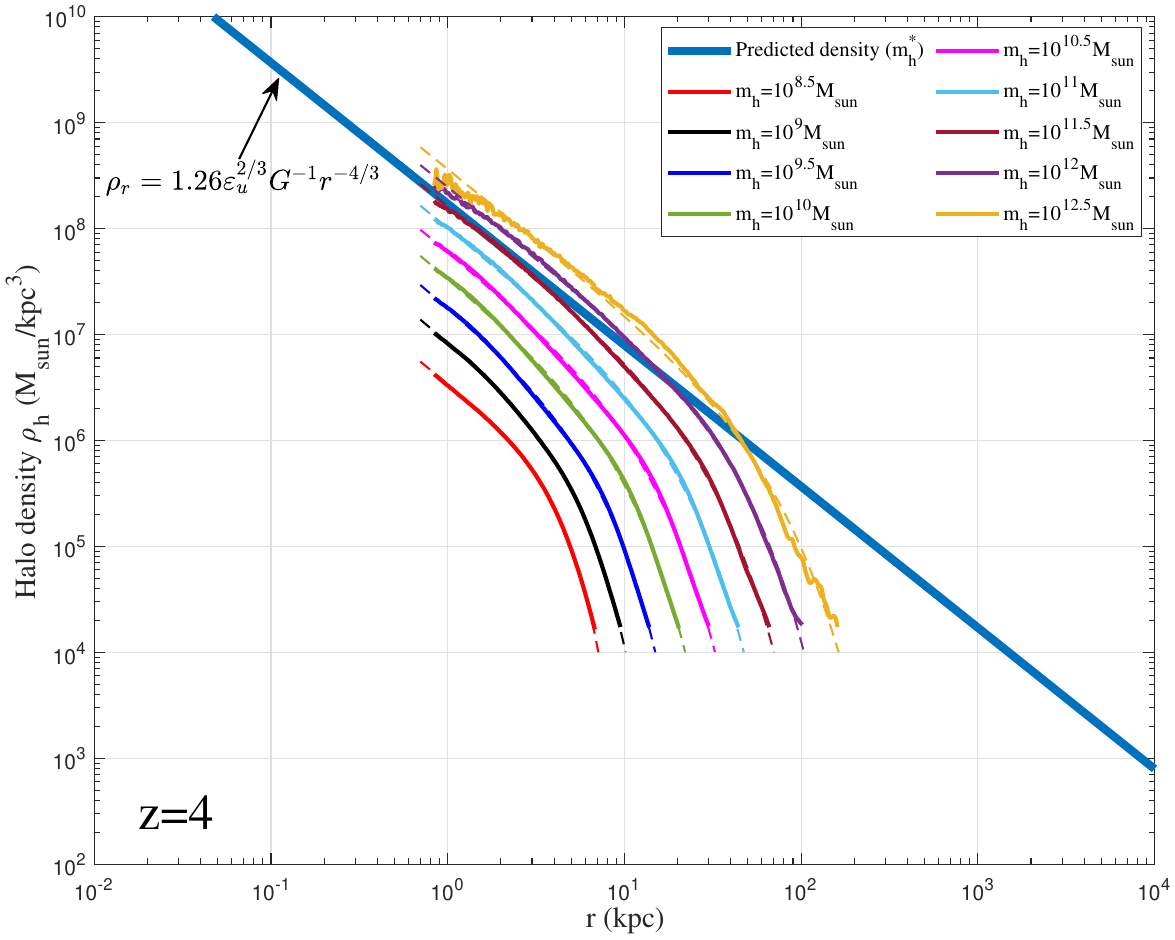}
\caption{Halo density profiles for different halo mass $m_h$ at $z=4$ (solid lines). The predicted scaling law (Eq. \eqref{eq:28}) for halo density is presented as the solid blue line. for comparison, the double-$\gamma$ density model (Eq. \eqref{eq:29}) was also plotted as dashed lines. Model fits better for halo density at higher redshift. The asymptotic density slope $-4/3$ at small $r$ can be identified.}
\label{fig:5-2}
\end{figure}

\begin{figure}
\includegraphics*[width=\columnwidth]{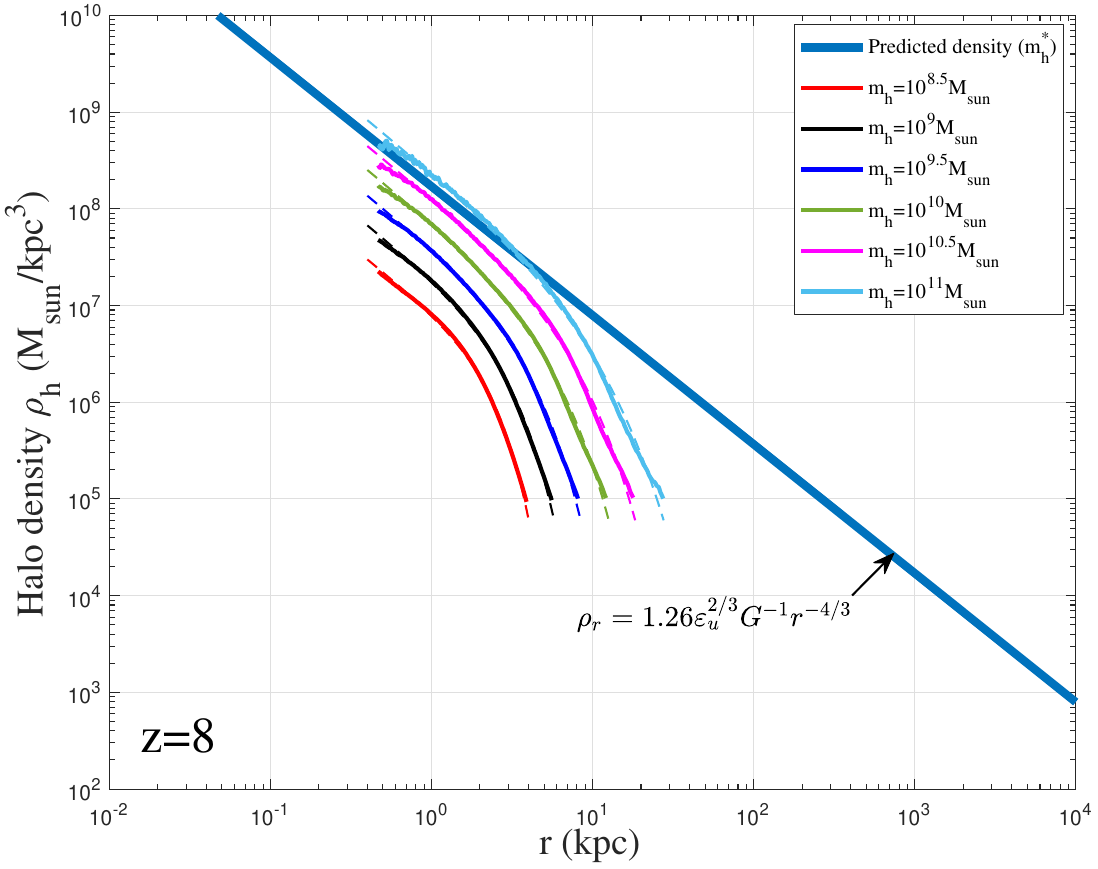}
\caption{Halo density profiles for different halo mass $m_h$ at $z=8$ (solid lines). The predicted scaling law (Eq. \eqref{eq:28}) for halo density is presented as the solid blue line. The double-$\gamma$ density model (Eq. \eqref{eq:29}) was also plotted as dashed lines. The asymptotic density slope $-4/3$ at small $r$ can be identified.}
\label{fig:5-3}
\end{figure}

\begin{figure}
\includegraphics*[width=\columnwidth]{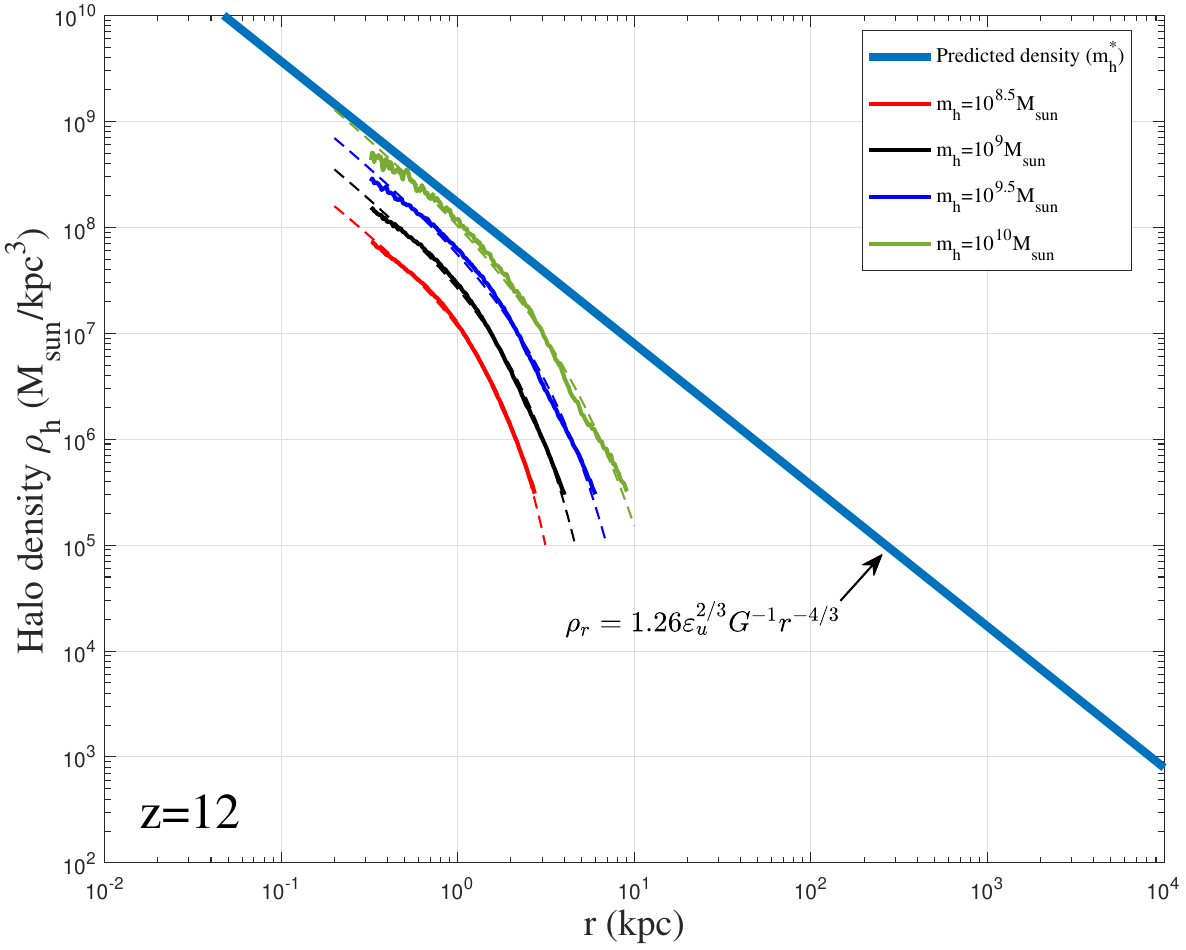}
\caption{Halo density profiles for different halo mass $m_h$ at $z=12$ (solid lines). The predicted scaling law (Eq. \eqref{eq:28}) for halo density is presented as the solid blue line. The double-$\gamma$ density model (Eq. \eqref{eq:29}) was also plotted as dashed lines for comparison.}
\label{fig:5-4}
\end{figure}

From this insight, assume $\gamma$ is unknown, we can predict the value of $\gamma$ as follows: Since the waiting time $\tau_g \propto \Phi(r)^{-1} \propto r^{-\gamma}$, halo density should scale as $\rho_r \propto r^{-2\gamma}$ from Eq. \eqref{eq:21}. The halo mass enclosed in $r$ scales as $m_r \propto \rho_r r^3 \propto r^{3-2\gamma}$. The local potential at $r$ should scale as $\Phi(r) \propto Gm_r/r \propto r^{3-2\gamma-1}$. The waiting time of particle at $r$ should satisfy Eq. \eqref{eq:18} that requires $3-2\gamma-1=\gamma$ such that $\gamma=2/3$ and the density slope $2\gamma=4/3$. It should be noted that the random walk theory for halo structure formation confirms the -4/3 law ($\rho_r\propto r^{-4/3}$) predicted by the energy cascade theory in Eq. \eqref{eq:28}. Predictions are tested against simulations in Figs. \ref{fig:5-1} to \ref{fig:5-4}. Similar to halo mass function (Eq. \eqref{eq:12}), the exponent $\gamma$ can be different in two different ranges, i.e. the power law below the scale radius $r_s$ and the exponential decay above $r_s$. Using two different $\gamma$ for $r$-dependence of waiting time $\tau_g(r) \propto r^{-\gamma}$, i.e. $\gamma_1$ and $\gamma_2$ for two different ranges, based on the single-$\gamma$ distribution in Eq. \eqref{eq:21}, the double-$\gamma$ distribution reads 
\begin{equation}
\label{eq:22} 
P_{r} \left(r,t\right) = \frac{(2-2\gamma_2)^{\frac{2\gamma_1-2-\gamma_2}{1-\gamma_2}} r^{-2\gamma_1} }{4\pi\left(D_0t\right)^{\frac{3-2\gamma_1}{2-2\gamma_2}}\Gamma\left(\frac{3-2\gamma_1}{2-2\gamma_2}\right)}\exp \left(-\frac{r^{2-2\gamma_2}}{4(1-\gamma_2)^2D_0t}\right).         
\end{equation} 
Introducing the conventional scale radius $r_s(t)$ where the logarithmic slope of $P_{r}(r,t)$ equals -2, we should have
\begin{equation}
\label{eq:23} 
4(1-\gamma_2)^2D_0t = \frac{2-2\gamma_2}{2-2\gamma_1}r_s^{2-2\gamma_2}.         
\end{equation} 
Substituting Eq. \eqref{eq:23} into Eq. \eqref{eq:22} and introducing a dimensionless spatial-temporal variable $x=r/r_s(t)$, distribution function reads
\begin{equation}
\label{eq:24} 
P_{r} \left(x\right) = \frac{(1-\gamma_2) x^{-2\gamma_1}}{2\pi\Gamma\left(\frac{3-2\gamma_1}{2-2\gamma_2}\right)\left(\frac{1-\gamma_2}{1-\gamma_1}\right)^{\frac{3-2\gamma_1}{2-2\gamma_2}}}\exp \left(-\frac{1-\gamma_1}{1-\gamma_2}x^{2-2\gamma_2}\right).         
\end{equation} 
Finally, the two parameter particle distribution function can be written as (with a similar form as mass function in Eq. \eqref{eq:13-2})
\begin{equation}
\label{eq:25} 
P_{r} \left(x\right) = \frac{\alpha \beta^{-(\frac{1}{\alpha}+\frac{1}{\beta})}}{4\pi\Gamma\left(\frac{1}{\alpha}+\frac{1}{\beta}\right)} x^{\frac{\alpha}{\beta}-2} \exp\left(-\frac{x^{\alpha}}{\beta}\right),      
\end{equation} 
where two dimensionless parameters $\alpha$ and $\beta$ are
\begin{equation}
\label{eq:26} 
\alpha = 2-2\gamma_2 \quad \textrm{and} \quad \beta = \frac{1-\gamma_2}{1-\gamma_1}.
\end{equation} 

The time variation of the distribution function is absorbed into the scale radius $r_s(t)$. The double-$\gamma$ distribution function reduces to the Einasto profile with $\alpha=2\beta$. The cumulative distribution in spherical coordinate can be easily obtained as,
\begin{equation}
\label{eq:27} 
\int_{0}^{x} P_{r} \left(y\right) 4\pi {y}^2 d{y} = 1-\frac{\Gamma\left(\frac{1}{\alpha}+\frac{1}{\beta},\frac{x^{\alpha}}
{\beta}\right)}{\Gamma\left(\frac{1}{\alpha}+\frac{1}{\beta}\right)},    
\end{equation} 
where $\Gamma(x,y)$ is an upper incomplete gamma function. 

So far we provide physical interpretation and a possible theory for halo density. The general density profile can be finally written as 
\begin{equation}
\label{eq:27-2} 
\rho_h(r,t)=\rho_s(t) \frac{P_r(x)}{P_r(1)} = \rho_s(t) x^{\frac{\alpha}{\beta}-2} \exp\left(\frac{1}{\beta}\left(1-{x^{\alpha}}\right)\right),
\end{equation}
where $\rho_s(t)$ is the density at scale radius $r_s$. Simulated haloes were found to have different density slopes in different simulations as discussed in Section \ref{sec:1}. This might be due to the different radial flow and mass accretion rate in these haloes, whose density profile can be modelled by the general solution in Eq. \eqref{eq:27-2} \citep{Xu:2023-Universal-scaling-laws-and-density-slope}. 

On small scale, virialized haloes are incompressible with vanishing (proper) radial flow \citep{Xu:2023-On-the-statistical-theory-of-self-gravitating}. For fully virialized haloes with vanishing radial flow, we would expect -4/3 law for inner density with $2\gamma_1=4/3$, which is consistent with the limiting density slope in Eq. \eqref{eq:28}. Combining Eq. \eqref{eq:27-2} with $\alpha/\beta=2/3$ leads to density profile that is consistent with the prediction from energy cascade in Eq. \eqref{eq:28},
\begin{equation}
\label{eq:29} 
\rho_h(r,t) = A_r\varepsilon_u^{2/3}G^{-1}r_s^{-4/3}\left(\frac{r}{r_s}\right)^{-4/3}\exp\left[-\frac{1}{\beta_r}\left(\frac{r}{r_s}\right)^{2\beta_r/3}\right].
\end{equation}
The small scale permanence for halo density in Fig. \ref{fig:4-3} becomes
\begin{equation}
\label{eq:29-1} 
\rho_h(r,t) \equiv \rho_h(r) =A_r \varepsilon_{u}^{2/3}G^{-1}r^{-4/3} \quad \textrm{for} \quad r\rightarrow 0,
\end{equation}
where $A_r$ is an amplitude parameter of halo density, $\beta_r=\beta$ is a shape parameter of density profile, and $r_s$ is the scale radius. 

To validate the proposed density model in Eq. \eqref{eq:29}, spherical averaged density profile was first obtained for all haloes with given mass in a range of $10^{\pm\Delta}m_h$ at different redshifts $z$. Next, we obtained the average halo density profile for all haloes in the same range at same redshift. The radial flow in these haloes might be cancelled out after this averaging such that the averaged halo density can be better described by Eq. \eqref{eq:29} with an inner slope of $2\gamma_1=4/3$. 

\begin{figure}
\includegraphics*[width=\columnwidth]{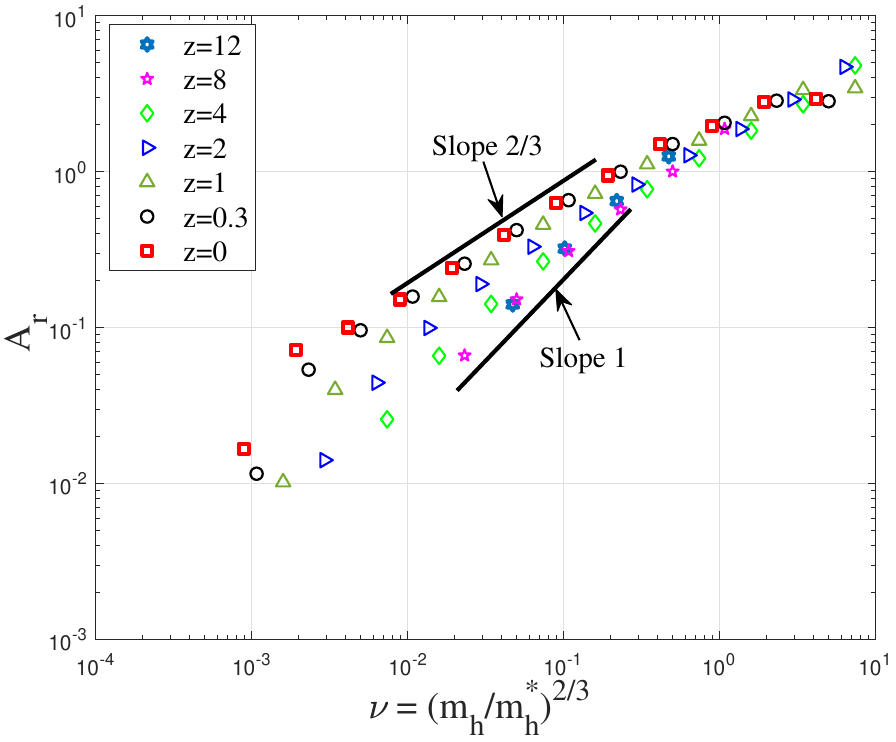}
\caption{The variation of amplitude parameter $A_r$ for halo density with the dimensionless parameter $\nu$ at different redshifts $z$. In principle, $A_r$ increases with halo mass $m_h$. This is related to the waiting time $\tau_g \propto m_h^{-\lambda}$.}
\label{fig:14}
\end{figure}

\begin{figure}
\includegraphics*[width=\columnwidth]{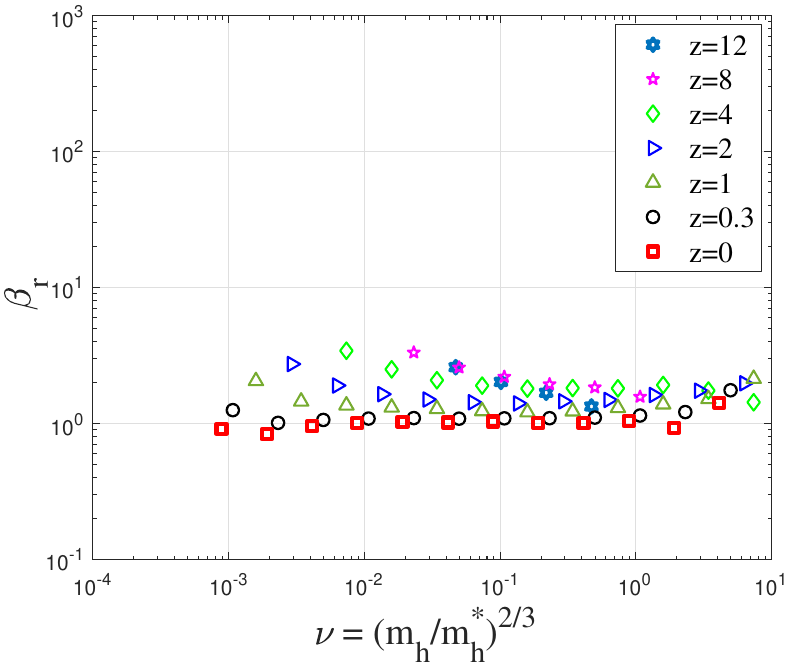}
\caption{The variation of shape parameter $\beta_r$ for halo density with $\nu$ at different redshifts $z$. The shape parameter $\beta_r$ varies in a small range between 1 and 3 and slightly decreases with halo mass $m_h$.}
\label{fig:15}
\end{figure}

\begin{figure}
\includegraphics*[width=\columnwidth]{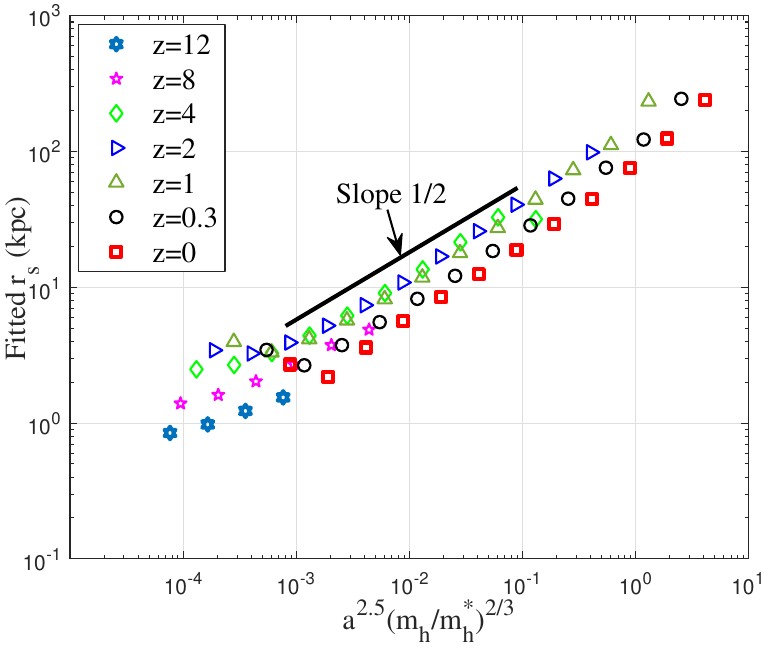}
\caption{The variation of scale radius $r_s$ for halo density with $\nu$ at different redshifts $z$. The scale radius increases with $\nu$ as $r_s \propto \nu^{1/2}$.}
\label{fig:16}
\end{figure}

Figures \ref{fig:5-1} to \ref{fig:5-4} present the halo density profiles of different halo mass $m_h$ at different redshifts $z$ from Illustris dark matter only simulations: Illustris-1-Dark (solid lines), where $\Delta$ is selected to be 0.1. The double-$\gamma$ density model (Eq. \eqref{eq:29}) was also used to fit all haloes and plotted as dashed lines in these figures. The best-fit model parameters $A_r$, $\beta_r$ and $r_s$ can be obtained for different halo mass $m_h$ and redshifts $z$ (as presented in Figs. \ref{fig:14} to \ref{fig:16}). The double-$\gamma$ density model provides a reasonably well fit to all haloes at all redshifts, with slightly better fit at higher redshift in a matter-dominant universe.

Figure \ref{fig:14} presents the variation of amplitude parameter $A_r$ with the dimensionless parameter $\nu$ defined in Eq. \eqref{eq:13-3}. As expected, the amplitude parameter $A_r \propto \nu^{2/3}$ increases with halo mass $m_h$ at fixed redshift or decreases with time at fixed mass $m_h$. The mass cascade across haloes is accompanied by a simultaneous energy cascade across haloes. The rate of cascade is independent of mass scale for group of haloes of the same mass. For individual haloes with mass $m_h<m_h^*$, the rate of energy cascade $\varepsilon$ in these haloes is smaller due to the longer waiting time $\tau_g \propto m_h^{-\lambda}$. The effective rate of energy cascade $\varepsilon$ in individual haloes is inversely proportional to $\tau_g$,
\begin{equation}
\label{eq:29-2} 
\varepsilon(m_h,a) = \left({m_h}/{m_h^*}\right)^{\lambda} \varepsilon_u = \nu^{3\lambda/2}\varepsilon_u.
\end{equation}
Therefore, the halo density $\rho_h \propto \varepsilon^{2/3} G^{-1} r^{-4/3}\propto m_h^{2\lambda/3}$ (see Eq. \eqref{eq:28}) such that the amplitude parameter $A_r \propto \nu^{\lambda}$, as shown in Fig. \ref{fig:14}. With $\lambda=2/3$, halo density scales with halo mass as $\rho_h \propto m_h^{4/9}$ at a given position $r$. 

Figure \ref{fig:15} presents the variation of shape parameter $\beta_r$ with $\nu$. The shape parameter $\beta_r$ is relatively independent of parameter $\nu$ at low redshift $z$. It varies in a small range between 1 and 3 and slightly decreases with halo mass $m_h$, which corresponds to a range of $\gamma_2=2/3$ for large haloes and $\gamma_2=0$ for small haloes with $\gamma_1=2/3$ (see Eq. \eqref{eq:26}). In the range $r>r_s$, the potential $\Phi$ is relatively independent of $r$ due to exponential decay of density. Therefore, the waiting time becomes less dependent on $r$ in this range with $\gamma_2\le\gamma_1$. Table \ref{tab:2} lists relevant values of $\lambda$ and $\gamma$ in different ranges. 

\begin{table}
\begin{center}
\centering
\scriptsize
\caption{Halo parameters $\lambda$ and $\gamma$ from theory and simulation}
\begin{tabular}{cccccc}
\hline\hline
Mass range  &  Scale Range  & \makecell{$\lambda$\\(pred.)} & \makecell{$\lambda$\\(simu.)}&\makecell{$\gamma$\\(pred.)}&\makecell{$\gamma$\\(simu.)} \\ 
\makecell{Small haloes \\ $m_h<m_h^*$}   &  \makecell{Core region\\ $r<r_s$}   & $\lambda_1=2/3$  & $\lambda_1=0.856$ & $\gamma_1=2/3$ & $\gamma_1=2/3$\\
\makecell{Small haloes \\ $m_h<m_h^*$}   &  \makecell{Outer region\\ $r>r_s$}  & $\lambda_1=2/3$  & $\lambda_1=0.856$ & $\gamma_2=2/3$ & $\gamma_2=0$\\
\makecell{Large haloes \\ $m_h>m_h^*$}   &  \makecell{Core region\\ $r<r_s$}   & $\lambda_2=2/3$  & $\lambda_2=0.605$ & $\gamma_1=2/3$ & $\gamma_1=2/3$\\ 
\makecell{Large haloes \\ $m_h>m_h^*$}   &  \makecell{Outer region \\ $r>r_s$} & $\lambda_2=2/3$  & $\lambda_2=0.605$ & $\gamma_2=2/3$ & $\gamma_2=2/3$ \\ 
\hline\hline
\label{tab:2} 
\end{tabular}
\end{center}
\end{table}

Figure \ref{fig:16} presents the variation of the best fitted scale radius $r_s$ with $\nu$ at different redshifts $z$, where $r_s$ increases with $\nu$ with an approximate scaling of $r_s \propto \nu^{1/2}$. In summary, the amplitude parameter $A_r$ is related to the rate of cascade $\varepsilon$ in haloes (Eq. \eqref{eq:29-2}), while the shape parameter $\beta_r$ is related to the parameter $\gamma$ (Eq. \eqref{eq:26}), i.e. the position dependence of waiting time $\tau_g\propto r^{-\gamma}$.

\rev{
It would be also interesting to compare the density profile obtained in this work with the Einasto and NFW profiles. Figure \ref{fig:17} presents the comparison for small ($10^{8.5}M_{\odot}$ ) and large haloes ($10^{13}M_{\odot}$) at redshift $z=0$ (haloes in Fig. \ref{fig:5-1}). These density profiles include: 1) the general double-$\gamma$ profile in Eq. \eqref{eq:27-2} with $\alpha$ and $\beta$ being independent; 2) the Einasto profile with $\alpha=2\beta$ in Eq. \eqref{eq:27-2}; 3) the double-$\gamma$ profile with $\alpha=2\beta/3$ in Eq. \eqref{eq:27-2} (or Eq. \eqref{eq:29}) for fully virialized haloes; and 4) the standard NFW profile. Bottom plots present the relative errors between these density profiles and simulation results. As expected, the general double-$\gamma$ profile provides the best fit of simulated halo density, compared to NFW profile. The double-$\gamma$ profile with $\alpha=2\beta/3$ (Eq. \eqref{eq:29}) provides a slight better fit than Einasto profile for small haloes, and a much better fit for large haloes. 

\begin{figure}
\includegraphics*[width=\columnwidth]{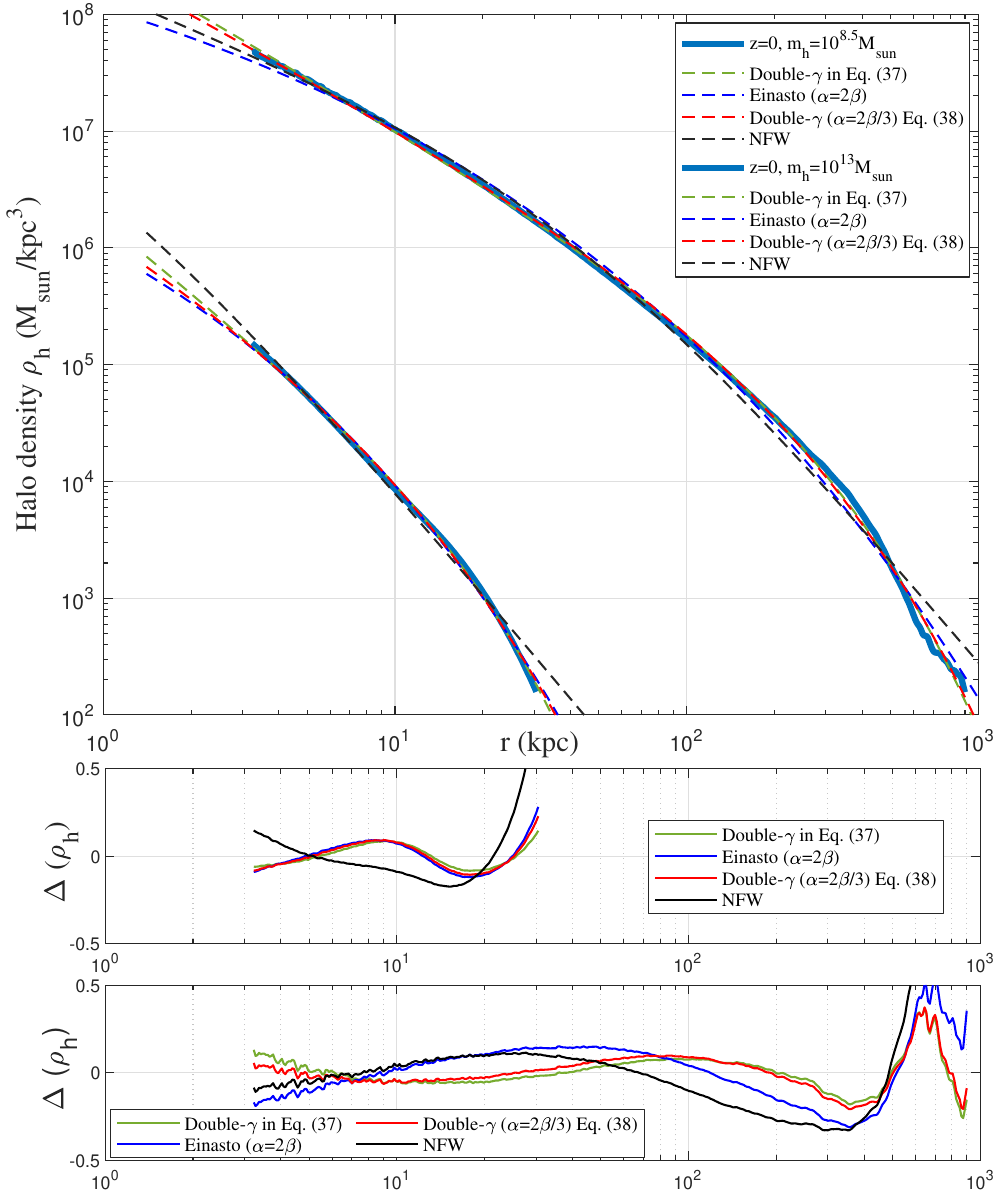}
\caption{The comparison between different density profiles that fit to haloes with a mass of $10^{8.5}M_{\odot}$ and $10^{13}M_{\odot}$ at redshift $z=0$. These density profiles include: 1) the general double-$\gamma$ profile in Eq. \eqref{eq:27-2} with $\alpha$ and $\beta$ being independent (green); 2) the Einasto profile with $\alpha=2\beta$ in Eq. \eqref{eq:27-2} (blue); 3) the double-$\gamma$ profile with $\alpha=2\beta/3$ in Eq. \eqref{eq:27-2} (or Eq. \eqref{eq:29}) (red); 4) the standard NFW profile (black). The bottom plots present the relative errors between these density profiles and simulation results. Double-$\gamma$ profiles provide better fit of simulated halo density.}
\label{fig:17}
\end{figure}

Finally, additional tests for different halo definitions and cosmologies should be very helpful to include data from simulations other than Illustris series. In this case, parameters in halo mass function and density models (Eqs. \eqref{eq:13} and \eqref{eq:27-2}) need to be fitted for different cosmologies. From this study, we can find how model parameters (halo parameters $\lambda$ and $\gamma$) vary with different cosmologies, which will require extensive work in future study. Here a quick test of double-$\gamma$ density for some simulated haloes in the literature was presented. Figure \ref{fig:2} provides the best fit by the general model in Eq. \eqref{eq:27-2} for these simulated haloes. Since the analytically derived double-$\gamma$ profile reduces to Einasto profile for $\alpha=2\beta/3$, the general double-$\gamma$ profile is expected to provide a better fit than Einasto profile for all simulated haloes.

\begin{figure}
\includegraphics*[width=\columnwidth]{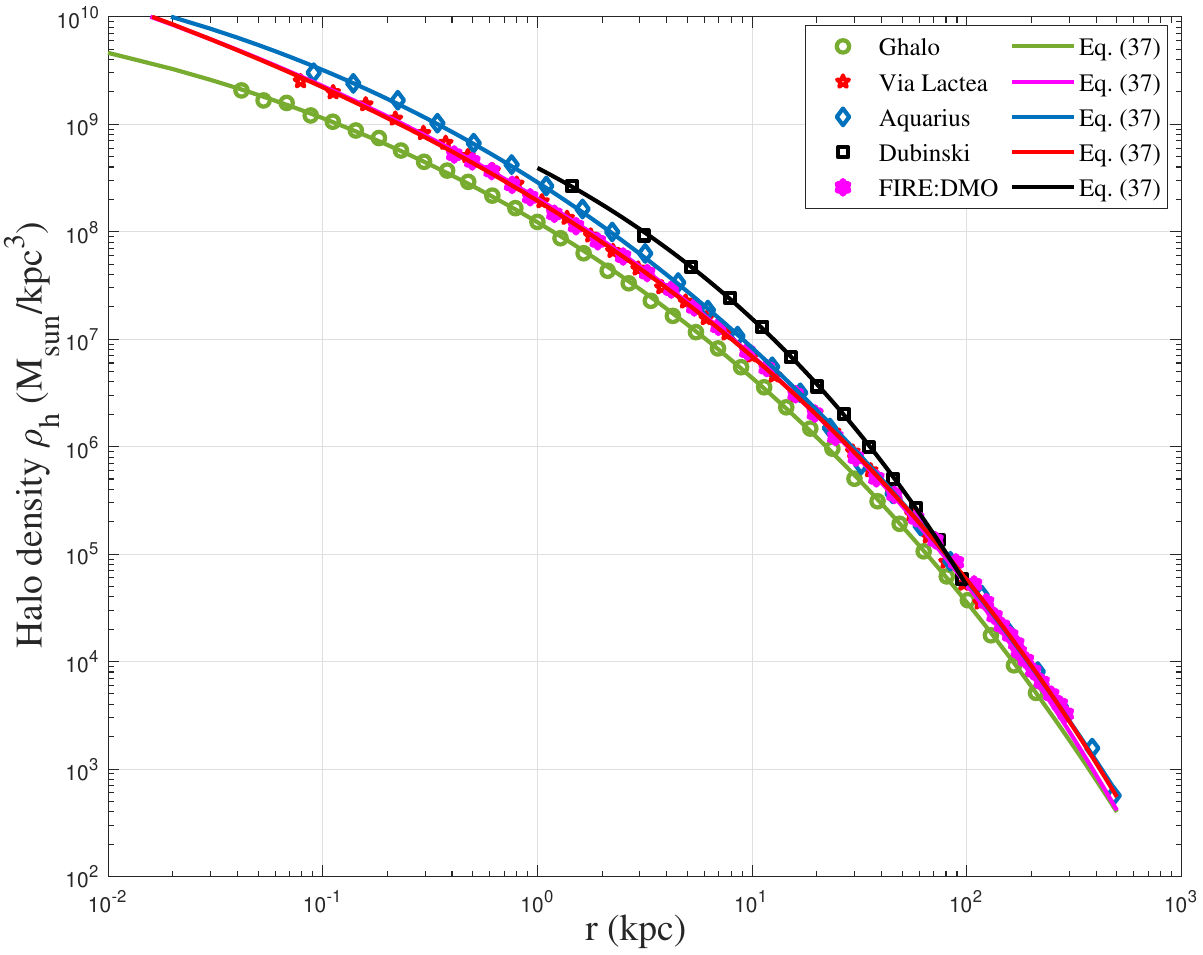}
\caption{Some halo density profiles for simulated haloes: 1) Ghalo \citep{Stadel:2009-Quantifying-the-heart-of-darkness}; 2) Via Lactea \citep{Diemand:2008-Clumps-and-streams-in-the-local}; 3) Aquarius \citep{Springel:2008-The-Aquarius-Project}; 4) Dubinski \citep{Dubinski:1991-The-Structure-of-Cold-Dark}; 5) FIRE:DMO \citep{McKeown:2022-Amplified-J-factors-in-the-Galactic-Centre}. The general double-$\gamma$ density model (Eq. \eqref{eq:27-2}) was also used to fit all simulated haloes for the entire range of $r$.}
\label{fig:2}
\end{figure}
}

\section{Conclusion}
\label{concl_sec}
In this paper, a simple theory was presented for halo mass function and density profile. The small scale permanence is proposed for halo group mass $m_g$ and halo density profile $\rho_h$ due to scale-independent rate of mass and energy cascade (Figs. \ref{fig:4-4} and \ref{fig:4-3}). Both halo mass function and halo density profile can be analytically derived based on this simple theory. The position-dependent waiting time $\tau_g\propto m_h^{-\lambda}$ leads to an analytical mass function modelled by a stretched Gaussian with a power-law behavior on small scale and exponential decay on large scale (Eq. \eqref{eq:10}). This can be further improved by considering two different values of $\lambda$ in propagation and deposition ranges, i.e. a double-$\lambda$ mass function in Eq. \eqref{eq:13}. Similarly, a double-$\gamma$ halo density profile is proposed based on the particle random walk in 3D space with a position-dependent waiting time $\tau_g\propto r^{-\gamma}$ (Eq. \eqref{eq:27-2}). The predicted value of $\gamma=2/3$ leads to a cuspy density profile with an inner slope of -4/3, consistent with the energy cascade theory (Eq. \eqref{eq:28}). The Press-Schechter mass function and Einasto profile are just special cases of the proposed model. Models were compared and validated against the Illustris simulations. Future work will involve additional tests for proposed models in different cosmologies. 


\section*{Data Availability}
Two datasets for this article, i.e. a halo-based and correlation-based statistics of dark matter flow, are available on Zenodo at \href{http://doi.org/10.5281/zenodo.6541230}{http://doi.org/10.5281/zenodo.6541230} \citep{Xu:2022-Dark_matter-flow-dataset-part1,Xu:2022-Dark_matter-flow-dataset-part2}, along with the accompanying presentation "A comparative study of dark matter flow \& hydrodynamic turbulence and its applications" \citep{Xu:2022-Dark_matter-flow-and-hydrodynamic-turbulence-presentation}.

\section*{Acknowledgements}
This research was supported by Laboratory Directed Research and Development at Pacific Northwest National Laboratory (PNNL). PNNL is a multiprogram national laboratory operated for the U.S. Department of Energy (DOE) by Battelle Memorial Institute under Contract no. DE-AC05-76RL01830. We acknowledge helpful discussions with Prof. Ethan Vishniac, Prof. Tom Abel, Prof. Tom Quinn, and Prof. Frank van den Bosch.

\bibliographystyle{Papers}
\bibliography{Papers}

\label{lastpage}
\end{document}